\documentclass[aip,jcp,reprint]{revtex4-1}
\pdfoutput=1
\usepackage[utf8]{inputenc}
\usepackage[unicode]{hyperref}
\usepackage{amsmath,amsfonts,amsthm}
\usepackage{graphicx,tabularx}
\usepackage{algpseudocode,enumitem}
\usepackage{amsmath}
\usepackage{amsfonts}
\usepackage{amssymb}
\usepackage{graphicx}
\usepackage{xcolor}
\hypersetup{pdfauthor={Miroslav Šulc, Henar Hernández, Todd J. Martínez, and Jiří Vaníček},pdftitle={Relation of exact Gaussian basis methods to the dephasing representation:
Theory and application to time-resolved electronic spectra},pdfkeywords={time-resolved stimulated emission, time correlation function, dephasing
representation, Gaussian basis method, ultrafast electronic spectrum}}
\providecommand{\U}[1]{\protect\rule{.1in}{.1in}}

\graphicspath{{./figures/}}
\begin{document}
\title{ Relation of exact Gaussian basis methods to the dephasing representation:
Theory and application to time-resolved electronic spectra }
\author{Miroslav \v{S}ulc}
\affiliation{Laboratory of Theoretical Physical Chemistry, Institut des Sciences et
Ing\'{e}nierie Chimiques, Ecole Polytechnique F\'{e}d\'{e}rale de Lausanne
(EPFL), CH-1015 Lausanne, Switzerland}
\author{Henar Hern\'{a}ndez}
\affiliation{Laboratory of Theoretical Physical Chemistry, Institut des Sciences et
Ing\'{e}nierie Chimiques, Ecole Polytechnique F\'{e}d\'{e}rale de Lausanne
(EPFL), CH-1015 Lausanne, Switzerland}
\affiliation{Departamento de F\'{\i}sica y Mec\'{a}nica, Universidad Polit\'{e}cnica, 28040
Madrid, Spain}
\author{Todd J. Mart\'{\i}nez}
\affiliation{Department of Chemistry, Stanford University, Stanford, CA 94305-5080}
\author{Ji\v{r}\'{\i} Van\'{\i}\v{c}ek}
\email{jiri.vanicek@epfl.ch}
\affiliation{Laboratory of Theoretical Physical Chemistry, Institut des Sciences et
Ing\'{e}nierie Chimiques, Ecole Polytechnique F\'{e}d\'{e}rale de Lausanne
(EPFL), CH-1015 Lausanne, Switzerland}

\begin{abstract}
We recently showed that the Dephasing Representation (DR)\ provides an
efficient tool for computing ultrafast electronic spectra and that further
acceleration is possible with cellularization [M.~\v{S}ulc and
J.~Van\'{\i}\v{c}ek, Mol.~Phys.\ \textbf{110}, 945 (2012)]. Here we focus on
increasing the accuracy of this approximation by first implementing an exact
Gaussian basis method, which benefits from the accuracy of quantum dynamics
and efficiency of classical dynamics. Starting from this exact method, the DR
is derived together with ten other methods for computing time-resolved spectra
with intermediate accuracy and efficiency. These methods include the Gaussian
DR,\ an exact generalization of the DR, in which trajectories are replaced by
communicating\ frozen Gaussian basis functions evolving classically with an
average Hamiltonian. The newly obtained methods are tested numerically on time
correlation functions and time-resolved stimulated emission spectra in the
harmonic potential, pyrazine $S_{0}/S_{1}$ model, and quartic oscillator.
Numerical results confirm that both the Gaussian basis method and the Gaussian
DR\ increase the accuracy of the DR. Surprisingly, in chaotic systems the
Gaussian DR can outperform the presumably more accurate Gaussian basis method,
in which the two bases are evolved separately.

\end{abstract}
\date{\today}
\maketitle

\section{Introduction}

High time resolution (such as $10^{-15}\,$s) is essential for understanding
many quantum dynamical processes in chemical physics and has been the main
challenge of ultrafast spectroscopy for over two
decades.\cite{Bisgaard2009,*Bressler2009,*Carbone2009} In theoretical studies,
in contrast, short time scales should simplify matters by requiring shorter
simulations. Still, solving the time-dependent Schr\"{o}dinger equation (TDSE)
is challenging even for short times due to the exponential scaling with
dimensionality. In practice, one must seek a~compromise between accuracy and
computational efficiency, which is provided, e.g., by
semiclassical\cite{miller:2001,Herman_94,*ThossWang_04,*Kay_05} or
time-dependent
finite-basis\cite{book_MCTDH,*Meyer:1990,Burghardt_99,Martinez_ACP_02}
methods. Both approaches benefit from the ultrafast character of the dynamics
not only thanks to a~lower computational cost, but also because their accuracy
deteriorates at longer times. Among these methods, semiclassical initial value
representation\cite{Tatchen_09,*Ceotto_JCP_09,*Ceotto_JCP_2013} and methods
employing Gaussian
bases\cite{Martinez_ACP_02,Martinez_FD_11,*Worth.Burghardt_FD_2004,*Worth.Lasorne_MolPhys_2008,*Saita.DVS_JCP_2012}
were employed successfully for \textquotedblleft direct\textquotedblright%
\ dynamics in which the electronic structure is evaluated on the fly.

In this paper, we propose an, in-principle, exact Gaussian basis method that
generalizes and increases the accuracy of the \textit{Dephasing
Representation}\cite{Vanicek_04,Vanicek_06} (DR), an efficient~semiclassical
approximation particularly fitted for calculations of time-resolved electronic
spectra.\cite{Wehrle_11,Sulc_MolPhys_12} In electronic spectroscopy, the DR
and closely related approximations are known as \textit{phase averaging}%
,\cite{Mukamel_82,*book_Mukamel} Wigner-averaged classical limit, or
linearized semiclassical initial value representation\cite{Shi_05,Geva:2009}
(LSC-IVR, in the generalized sense\nocite{Miller_LSCIVR}\cite{note2}), and
have been used by several
authors.\cite{Li_96,*Egorov_98,*Egorov_99,Shi_05,Geva:2009,Shemetulskis_92,*Rost_95}%

Although the original formulation of the DR pertains to a~single electronic
potential energy surface, a~generalization to multiple surfaces
exists.\cite{Zimmermann_12,Zimmermann_12b} The DR has many other applications:
e.g., in inelastic neutron scattering,\cite{Petitjean_07} as a~measure of the
dynamical importance of diabatic,\cite{Zimmermann_10}
nonadiabatic,\cite{Zimmermann_12} or spin-orbit
couplings,\cite{Zimmermann_12b} and as a~measure of the accuracy of quantum
molecular dynamics on an approximate potential energy
surface.\cite{Li_09,*Zimmermann_IJQC_10} In the field of quantum chaos, DR
successfully describes the local density of states and transition from the
Fermi-Golden-Rule to the Lyapunov regime of fidelity
decay.\cite{Wang_05,*Ares_09,*Wisniacki_10,*Wisniacki_11}

The most attractive feature of the DR is its efficiency, which is, as in the
forward-backward semiclassical dynamics of Makri and
co-workers,\cite{Thompson_Makri:1999,*Kuhn_Makri:1999} partially due to the
reduction of the sign problem. Motivated by numerical comparisons with other
semiclassical methods,\cite{Wehrle_11} we recently proved
analytically\cite{Mollica_PRL_11} that the number of trajectories required for
convergence of the DR is independent of the system's dimensionality,
Hamiltonian, or total evolution time. Inspired by Heller's cellular
dynamics,\cite{Heller_91} we have further increased computational efficiency
of the DR by formulating a~\textit{cellular DR.}\cite{Sulc_MolPhys_12}

Unlike its efficiency, the accuracy of the DR is not always sufficient. The DR
is exact in displaced harmonic oscillators\cite{Mukamel_82,*book_Mukamel} and
often accurate in chaotic systems,\cite{Vanicek_04,Vanicek_06} but due to its
perturbative nature, the DR breaks down, e.g., in harmonic oscillators with
significantly different force constants. Whereas Zambrano and Ozorio de
Almeida proposed to correct DR with a~prefactor,\cite{Zambrano_PRE_11} in this
paper the accuracy of DR\ is increased with a~Gaussian basis approach.

Since any quantum dynamics can be performed in a~Gaussian basis, methods
employing Gaussian bases should be useful also for time-resolved spectroscopy.
Any basis-set approach is, in principle, exact; the only inexactness\ stems
from the incompleteness of the basis. As a~result, the goal is to find the
smallest basis giving sufficiently converged result. A~useful way to reduce
basis size is to employ time-dependent bases that explore all dynamically
important regions of phase space. Such an approach has been explored
extensively in the Multi-Configuration Time-Dependent Hartree
(MCTDH),\cite{book_MCTDH,*Meyer:1990} Gaussian
MCTDH,\cite{Burghardt_99,Burghardt:2008} and Multiple
Spawning\cite{Martinez_ACP_02,Martinez_JCP_96,*Martinez_JPC_96} methods. Here
we propose two exact methods that are closely related to
these\cite{book_MCTDH,*Meyer:1990,Burghardt_99,Burghardt:2008,Martinez_ACP_02,Martinez_JCP_96,*Martinez_JPC_96}
and several
other\cite{Sawada_85,Metiu_CPL_85,*Metiu_JCP_86,Shalashilin_JCP_00,*Shalashilin_JCP_01a,*Shalashilin_JCP_01b,*Shalashilin_ChemPhys_04}
methods employing Gaussian bases, yet are specific to time-resolved
spectroscopy. One of the two methods, which we call the Gaussian Basis Method,
uses two bases evolving classically with two Hamiltonians corresponding to the
two electronic states. The other method employs a single basis evolved
classically with the average Hamiltonian. Because of its relation to the DR,
we call it the \textit{Gaussian DR}. Our results show that both the Gaussian
Basis Method and the Gaussian DR improve the accuracy of DR in time-resolved
stimulated emission spectra calculations in a~harmonic potential, pyrazine
$S_{0}/S_{1}$ model, and chaotic quartic oscillator.

Moreover, we show that the DR emerges naturally from the exact Gaussian Basis
Method by a sequence of three approximations: propagating the basis with the
average Hamiltonian (which gives the Gaussian DR), using independent
Gaussians, and assuming local approximation for the potential. Since the three
approximations may be taken in arbitrary order and the local approximation
relaxed to a~local harmonic approximation, we derive ten
intermediate\ approximations, potentially useful for future applications. We
observe a~remarkable property that using the average Hamiltonian for
propagating the basis, which is seemingly an approximation, can sometimes
outperform the original Gaussian Basis Method. This occurs particularly in
chaotic systems and parallels a~property of the semiclassical DR.

The rest of this paper is organized as follows: Section~\ref{sec:theory}
describes the central theoretical concepts. Section~\ref{sec:results} uses
methods from Sec.~\ref{sec:theory} to compute time correlation functions and
time-resolved stimulated emission spectra, and Section~\ref{sec:conclusions}
provides conclusions. An essential part of the paper is the Appendix
describing an efficient numerical implementation of the methods from
Sec.~\ref{sec:theory}.


\section{Theory\label{sec:theory}}

\subsection{Time-resolved stimulated emission: spectrum, time correlation
function, and Dephasing Representation}

To be specific, we restrict the discussion to time-resolved stimulated
emission (TRSE). Within the electric dipole approximation, time-dependent
perturbation theory, and ultrashort pulse approximation, this spectrum can be
computed as a~Fourier transform of the following correlation
function:\cite{Wehrle_11,Sulc_MolPhys_12}
\begin{align}
C_{\text{TRSE}}(t,\,\tau)  &  =E_{\text{pu}}^{2}\,E_{\text{pr}}\,{\mathrm{Tr}%
[}\hat{\rho}_{0}(T)\hat{\mu}_{01}\hat{U}_{1}(-t-\tau)\nonumber\\
&  \times\hat{\mu}_{10}\hat{U}_{0}(t)\hat{\mu}_{01}\hat{U}_{1}(\tau)\hat{\mu
}_{10}]. \label{eq:corr_func}%
\end{align}
Above, $E_{\text{pu}}$ and $E_{\text{pr}}$ are the amplitudes of the pump and
probe laser pulses, $\hat{\rho}_{0}(T)$ represents the nuclear density
operator in the electronic ground state at temperature $T$, $\hat{\mu}_{ij}$
is the transition dipole moment operator coupling electronic states $i$ and
$j$, $\tau$ stands for the time delay between the pump and probe pulses, and
$t$ is time after the probe pulse. Finally, $\hat{U}_{j}$ denotes the nuclear
quantum evolution operator
\begin{equation}
\hat{U}_{j}(t)=\exp(-i\hat{H}_{j}t/\hbar) \label{eq:evolution_op}%
\end{equation}
with Hamiltonian $\hat{H}_{j}=\hat{T}+\hat{V}_{j}$, where $\hat{T}$ is the
kinetic energy operator and $\hat{V}_{j}$\ is the $j$th potential energy
surface. In all expressions, the hat denotes operators in the Hilbert space of nuclei.

Within the Franck-Condon approximation and in the zero temperature limit,
correlation function~(\ref{eq:corr_func}) simplifies to
\begin{equation}
C_{\text{TRSE}}(t,\,\tau)=E_{\text{pu}}^{2}\,E_{\text{pr}}\lvert\mu_{10}%
\rvert^{4}f(t,\,\tau),
\end{equation}
where
\begin{align}
f(t,\,\tau)  &  :=\langle\psi_{1}(t,\,\tau)|\psi_{0}(t,\,\tau)\rangle
,\label{eq:fid_amp}\\
\lvert\psi_{j}(t,\,\tau)\rangle &  :=\hat{U}_{j}(t)\hat{U}_{1}(\tau)\lvert
\Psi_{\text{init}}\rangle, \label{eq:psi_J_t}%
\end{align}
is a~specific time correlation function and the initial state $\lvert
\Psi_{\text{init}}\rangle$ is generally the vibrational ground state of the
ground potential energy surface. The TRSE spectrum, given
by\cite{Pollard_Mathies:1990}
\[
\sigma_{\text{TRSE}}(\omega,\tau)\propto\omega\,E_{\text{pu}}^{2}%
\,E_{\text{pr}}\lvert\mu_{10}\rvert^{4}\sigma(\omega,\tau),
\]
is proportional to the wavepacket spectrum $\sigma$ obtained\cite{book_Tannor}
via the Fourier transform of $f$:
\begin{equation}
\sigma(\omega,\,\tau)=\operatorname{Re}\int_{0}^{\infty}\!\!\!d{t}%
\,\,f(t,\,\tau)\,e^{i{}{\omega{}t}}. \label{eq:sigma_spec}%
\end{equation}

Correlation function~(\ref{eq:fid_amp}) specific to the stimulated emission is
a~special case of a~more general concept of \textit{fidelity amplitude}%
,\cite{Gorin_06,Jacquod_09} defined as
\begin{equation}
f(t_{f})=\langle\Psi_{\text{init}}|\hat{U}_{\text{I}}(t_{f},0)^{-1}\hat
{U}_{\text{II}}(t_{f},0)\lvert\Psi_{\text{init}}\rangle,
\label{eq:fid_amp_gen}%
\end{equation}
where $\hat{U}_{J}(t_{f},0)$, $J=\,$I, II, is the time evolution operator for
a~time-dependent Hamiltonian $\hat{H}_{J}(\tilde{t})$:
\begin{equation}
\hat{U}_{J}(t_{f},0)=\mathcal{T}\exp\left[  -\frac{i}{\hbar}\int_{0}^{t_{f}%
}\!d{\tilde{t}}\,\hat{H}_{J}(\tilde{t})\right]  . \label{eq:evolution_op_gen}%
\end{equation}

Correlation function~(\ref{eq:fid_amp}) for TRSE is obtained from the general
fidelity amplitude~(\ref{eq:fid_amp_gen}) if the time-dependent Hamiltonians
$\hat{H}_{J}(t)$ in Eq.~(\ref{eq:evolution_op_gen}) are
\begin{align*}
\hat{H}_{\text{I}}({\tilde{t}})  &  =\phantom{\biggl\{}%
\begin{array}
[c]{lcl}%
\hat{H}_{1} & \text{ for } & 0\leq{\tilde{t}}\leq\tau+t,
\end{array}
\\
\hat{H}_{\text{II}}({\tilde{t}})  &  =\biggl\{%
\begin{array}
[c]{lcl}%
\hat{H}_{1} & \text{ for } & 0\leq{\tilde{t}}\leq\tau,\\[1.2ex]%
\hat{H}_{0} & \text{ for } & \tau\leq{\tilde{t}}\leq\tau+t.
\end{array}
\end{align*}
Besides applications in electronic
spectroscopy,\cite{Shemetulskis_92,*Rost_95,Li_96,*Egorov_98,*Egorov_99,Shi_05,Geva:2009}
correlation function~(\ref{eq:fid_amp_gen}) proved useful, e.g., in NMR spin
echo experiments\cite{Pastawski_00} and theories of quantum
computation,\cite{Gorin_06}
decoherence,\cite{Gorin_06,Jacquod_09,Cucchietti_03} and inelastic neutron
scattering.\cite{Petitjean_07} The fidelity amplitude was also used as
a~measure of the dynamical importance of diabatic,\cite{Zimmermann_10}
nonadiabatic,\cite{Zimmermann_12} or spin-orbit
couplings,\cite{Zimmermann_12b} and of the accuracy of quantum molecular
dynamics on an approximate potential energy
surface.\cite{Li_09,*Zimmermann_IJQC_10}

In practical calculations, correlation function~(\ref{eq:fid_amp_gen}) is
usually approximated, and DR provides an efficient semiclassical
approximation.\cite{Vanicek_04,Vanicek_06,Mukamel_82,*book_Mukamel,Shemetulskis_92,*Rost_95,Li_96,*Egorov_98,*Egorov_99,Shi_05,Geva:2009}
If we denote by $x^{t}:=(q^{t},\,p^{t})$ the phase-space coordinates at time
$t$ of a~point along a~classical trajectory of the \emph{average}
\cite{Mukamel_82,*book_Mukamel,Wehrle_11,Zambrano_PRE_11} Hamiltonian
$({H}_{\text{I}}+{H}_{\text{II}})/2$, the DR of fidelity
amplitude~(\ref{eq:fid_amp_gen}) is written as
\begin{equation}
f_{\text{DR}}(t,\,\tau)=h^{-D}\int d{x^{0}}\rho_{\text{W}}(x^{0}%
)\,e^{i{\Delta{}S({x^{0}}\!,t,\,\tau)}/\hbar}, \label{eq:fidelity_amp_DR}%
\end{equation}
with
\begin{equation}
\rho_{\text{W}}(q^{0}\!,\,p^{0})=\int\!d{s}\,\,e^{i{s}^{\mathrm{T}}%
\!\cdot{p^{0}/}\hbar}\langle{q}^{0}-s/2|\hat{\rho}_{\text{init}}%
|q^{0}+s/2\rangle. \label{eq:rhoW}%
\end{equation}
Here $D$ is the number of degrees of freedom, $\rho_{\text{W}}$ represents the
Wigner transform of the initial density operator $\hat{\rho}_{\text{init}%
}=\lvert\Psi_{\text{init}}\rangle\langle\Psi_{\text{init}}\rvert$, and
$\Delta{}S({x^{0}}\!,t,\,\tau)$ denotes the action due to the difference
$\Delta H:=H_{\text{II}}-H_{\text{I}}$ along trajectory $x^{t}$:
\begin{equation}
\Delta{}S({x^{0}}\!,t,\,\tau)=-\int_{0}^{t+\tau}\!\!d{\tilde{t}}\,\Delta
{}V(x^{\tilde{t}},{\tilde{t}}). \label{eq:dS}%
\end{equation}
For TRSE (\ref{eq:fid_amp}),
\begin{align}
\Delta{}V  &  =V_{\text{II}}-V_{\text{I}}\nonumber\\
&  =\biggl\{%
\begin{array}
[c]{lcl}%
0 & \text{ for } & 0\leq{\tilde{t}}\leq\tau,\\[1.2ex]%
V_{0}-V_{1} & \text{ for } & \tau\leq{\tilde{t}}\leq\tau+t.
\end{array}
\label{eq:delta_V}%
\end{align}
Throughout this paper, time dependence is denoted by $t$ as a~superscript or
argument in parentheses. Italics subscripts label either nuclear
($i\in\{1,\ldots,D\}$) or electronic ($i\in\{0,1\}$) degrees of freedom.
Vectors and matrices in the $D$-dimensional vector space of nuclei are denoted
by italics: e.g., $q$ or $p$. The inner product and contraction of tensors in
this space are denoted by $\cdot$,\ as in $q^{\text{T}}\cdot p$.

The DR (\ref{eq:fidelity_amp_DR}) can be derived\cite{Vanicek_04,Vanicek_06}
by linearization of the semiclassical propagator and improves on a~previous
method\cite{Vanicek_03} inspired by the semiclassical perturbation theory of
Miller and co-workers.\cite{MillerSmith_PRA_78,*Hubbard_JCP_83} Shi and
Geva\cite{Shi_05} derived the DR without invoking the semiclassical
propagator---by linearizing\cite{Poulsen_03,*Bonella_05,*Coker_11} the path
integral quantum propagator.

\subsection{\label{subsec:dynamics}Quantum dynamics and time correlation
functions in a~classically evolving Gaussian basis}

Since our main objective is to improve the accuracy of the
DR~(\ref{eq:fidelity_amp_DR}) by evaluating correlation
function~(\ref{eq:fid_amp}) without invoking the semiclassical perturbation
approximation, we solve the TDSE in a~classically evolving Gaussian basis.
Following Heller and co-workers,\cite{Heller_75,Davis_79,Heller_81} the
initial state and the state at time $t$ are expanded as%
\begin{align}
\lvert\Psi_{\text{init}}\rangle &  =\sum_{\alpha=1}^{N}c_{\alpha}\lvert
\phi_{\alpha}\rangle,\label{eq:Psi_init}\\
|\psi(t)\rangle &  =\sum_{\alpha=1}^{N}c_{\alpha}(t)|\phi_{\alpha}(t)\rangle,
\label{eq:psi}%
\end{align}
where the time-dependent basis $\{\phi_{\alpha}(t)\}_{\alpha=1}^{N}$ consists
of normalized Gaussians
\begin{equation}
|\phi_{\alpha}(t)\rangle=|\gamma_{\alpha},q_{\alpha}^{t},p_{\alpha}^{t}%
\rangle\label{eq:basis_fn}%
\end{equation}
with
\begin{align}
\langle q|\gamma,Q,P\rangle:=  &  \left(  \frac{\det\gamma}{\pi^{D}}\right)
^{\frac{1}{4}}\exp[i{P}^{\mathrm{T}}\!\cdot(q-Q)/\hbar\nonumber\\
&  -{(q-Q)}^{\mathrm{T}}\!\cdot{\gamma}\cdot{(q-Q)/2}]. \label{eq:gaussian}%
\end{align}
The real symmetric width matrix $\gamma$ in Eq.~(\ref{eq:gaussian}) is assumed
to be time-independent as in Heller's \textit{Frozen Gaussians Approximation}
(FGA).\cite{Heller_81} The time-dependent parameters $q_{\alpha}^{t}$ and
$p_{\alpha}^{t}$ in Eq.~(\ref{eq:basis_fn}) denote the position and momentum
of the center of the Gaussian state $\lvert\phi_{\alpha}\rangle$, which
evolves classically:%
\begin{equation}
\dot{x}_{\alpha}=\{x_{\alpha},H(x_{\alpha})\}. \label{eq:classical}%
\end{equation}
As in the MCTDH method,\cite{Beck_00} the evolution of the basis compensates
for its incompleteness. Above and throughout this paper, Greek subscripts,
such as $\alpha\in\{1,\ldots,N\}$, label basis functions or components of
vectors in the $N$-dimensional vector space spanned by $\{\phi_{\alpha}\}$.
Vectors and matrices in this space are denoted with the bold Roman font (e.g.,
$\mathbf{c}$ and $\mathbf{H}$ below); the inner product and contraction of
tensors are expressed by juxtaposition of matrices, as in $\mathbf{Hc}$.

Inserting ansatz~(\ref{eq:psi}) into the TDSE yields a~first-order
differential equation for the expansion coefficients:
\begin{equation}
{\mathbf{S}(t)}\,{\dot{\mathbf{c}}(t)}=-{\left[  \frac{i}{\hbar}%
\mathbf{H}(t)+\mathbf{D}(t)\right]  }\,{\mathbf{c}(t)}, \label{eq:working_eq}%
\end{equation}
where $\mathbf{S}(t)$ denotes the time-dependent overlap matrix
\begin{equation}
S_{\alpha\beta}(t):=\langle\phi_{\alpha}(t)|\phi_{\beta}(t)\rangle
\label{eq:mat_S_el}%
\end{equation}
and $\mathbf{H}(t)$ the Hamiltonian matrix
\begin{align}
H_{\alpha\beta}(t)  &  \equiv T_{\alpha\beta}(t)+V_{\alpha\beta}(t)\nonumber\\
&  =\langle\phi_{\alpha}(t)|\hat{T}|\phi_{\beta}(t)\rangle+\langle\phi
_{\alpha}(t)|\hat{V}(t)|\phi_{\beta}(t)\rangle. \label{eq:mat_H_el}%
\end{align}
The non-Hermitian time-de\-riv\-a\-tive\ matrix $\mathbf{D}$, defined as
\begin{equation}
D_{\alpha\beta}(t):=\langle\phi_{\alpha}(t)|\dot{\phi}_{\beta}(t)\rangle,
\label{eq:mat_D_el}%
\end{equation}
satisfies
\begin{equation}
\dot{\mathbf{S}}(t)=\mathbf{D}(t)^{\dagger}+\mathbf{D}(t).
\end{equation}
Our \textquotedblleft frozen\textquotedblright\ Gaussian basis functions
depend on time only via the classically evolving coordinates $q_{\alpha}%
^{t},\,p_{\alpha}^{t}$:
\begin{align}
\dot{\phi}_{\alpha}(q,\,t)  &  =\phi_{\alpha}(q,\,t)\left\{  {\dot{q}_{\alpha
}^{t}{}}^{\mathrm{T}}\!\cdot{\gamma_{\alpha}}\cdot{(q-q_{\alpha}^{t})}\right.
\nonumber\\
&  +\frac{i}{\hbar}\left.  \!\left[  {\dot{p}_{\alpha}^{t}{}}^{\mathrm{T}%
}\!\cdot(q-q_{\alpha}^{t})-{p_{\alpha}^{t}}^{\mathrm{T}}\!\cdot{{\dot{q}%
}_{\alpha}^{t}}\right]  \right\}  .
\end{align}
Analytical formulae for $\mathbf{S}$, $\mathbf{D}$, $\mathbf{T}$, and
$\mathbf{V}$ matrix elements are derived in Appendix~\ref{app:A}, while the
numerical implementation of the propagation algorithm is described in
Appendix~\ref{sec:numerics}.

Propagation equations~(\ref{eq:working_eq}) can be obtained also from the
Dirac-Frenkel variational principle applied to the ansatz~(\ref{eq:psi}) with
the coefficients $c_{\alpha}(t)$ playing the role of variational parameters.
Note, however, that the centers of individual Gaussians propagate along
classical trajectories of the original, classical Hamiltonian. This fact does
not follow from the variational principle but is enforced as in Heller's
work.\cite{Heller_75,Heller_81,Heller_06} Hence the propagation of the
Gaussians is uncoupled,\cite{book_MCTDH,*Meyer:1990} although--in contrast to
the Independent Gaussian Approximation\cite{Sawada_85} of
\citeauthor{Sawada_85}--the propagation of the expansion coefficients does
require communication between the Gaussians [see Eq.~(\ref{eq:working_eq})].
The Gaussian MCTDH\cite{Burghardt_99,Burghardt:2008,book_MCTDH,*Meyer:1990}
and Minimum Error\cite{Sawada_85,Metiu_CPL_85,*Metiu_JCP_86} methods, on the
other hand, treat both the expansion coefficients and all Gaussian parameters
variationally.\cite{Heller_JCP_76,Truhlar_JCP_84} The concept of classical
trajectories is modified also in the Coupled Coherent
States,\cite{Shalashilin_JCP_00,*Shalashilin_JCP_01a,*Shalashilin_JCP_01b,*Shalashilin_ChemPhys_04}
where individual Gaussians evolve according to the reordered Hamiltonian,
i.e., on a potential energy surface that is averaged over the width of the
Gaussian basis function. Finally, the propagation Eq.~(\ref{eq:working_eq}) is
a special case of the central equation of Multiple
Spawning,\cite{Martinez_ACP_02} which also considers couplings between
electronic states and allows the basis to change its size during dynamics.

Up to this point, the presentation applied to general quantum dynamics. Now we
describe how to use the Gaussian basis formalism to evaluate the correlation
function~(\ref{eq:fid_amp}) for TRSE. The initial state~(\ref{eq:Psi_init})
must be propagated with the two different propagators to obtain the two final
states~(\ref{eq:psi_J_t}). In analogy to Eq.~(\ref{eq:psi}), these final
states\cite{note1} are expanded as
\begin{equation}
\lvert\psi_{j}(t,\,\tau)\rangle=\sum_{\alpha=1}^{N}c_{j,\alpha}(t,\,\tau
)\lvert\phi_{j,\alpha}(t,\,\tau)\rangle, \label{eq:fid_12_psi_J}%
\end{equation}
where $c_{j,\alpha}(0,\,0)=c_{\alpha}$ and $\lvert\phi_{j,\alpha
}(0,\,0)\rangle=\lvert\phi_{\alpha}\rangle$. The two bases are propagated
classically with two separate equations (\ref{eq:classical}): one for
$x_{0,\alpha}$, the other for $x_{1,\alpha}$. Using Eq.~(\ref{eq:fid_12_psi_J}%
), fidelity amplitude~(\ref{eq:fid_amp}) becomes
\begin{equation}
f_{\text{GBM}}(t,\,\tau)=\mathbf{c}_{1}(t,\,\tau)^{\dagger}\,\mathbf{M}%
(t,\,\tau)\,\mathbf{c}_{0}(t,\,\tau) \label{eq:fid_12}%
\end{equation}
with
\[
M_{\alpha\beta}(t,\,\tau):=\langle\phi_{1,\alpha}(t,\,\tau)|\phi_{0,\beta
}(t,\,\tau)\rangle.
\]
Matrix elements $M_{\alpha\beta}$ differ from overlaps $S_{j,\alpha\beta
}:=\langle\phi_{j,\alpha}(t,\,\tau)|\phi_{j,\beta}(t,\,\tau)\rangle$ since
$\alpha$ and $\beta$ in $M_{\alpha\beta}$ denote basis functions evolved with
two different Hamiltonians. We refer to the method specified by
Eq.~(\ref{eq:fid_12}) simply as the \textit{Gaussian Basis Method} (GBM).

\subsection{\label{subsubsec:approx}Several approximations and derivation of
Dephasing Representation from the Gaussian Basis Method}

It is often necessary to treat the propagation Eq.~(\ref{eq:working_eq})
approximately. This is especially true in \textit{ab~initio\/} applications,
where the evaluation of the potential becomes expensive. Another reason is the
implicit matrix inversion in Eq.~(\ref{eq:working_eq}). We first discuss
approximations relevant for general quantum dynamics in a~Gaussian basis.

\textit{Independent Gaussians (IG). }This approximation avoids the inversion
problem as well as matrix multiplication by assuming that
\begin{equation}
S_{\alpha\beta}\approx\delta_{\alpha\beta},\label{eq:mat_S_delta}%
\end{equation}
which is justified if the basis is sparse enough so that different basis
functions have a~negligible overlap. As derived in Appendix~\ref{app:A}, if
Eq.~(\ref{eq:mat_S_delta}) is satisfied exactly then the $\mathbf{D}$ and
$\mathbf{H}=\mathbf{T}+\mathbf{V}$ matrices are diagonal. For $\mathbf{D}$ and
$\mathbf{T}$ matrices this statement follows directly from
Eqs.~(\ref{app:eq:mat_D}) and (\ref{app:eq:mat_T}). As for $\mathbf{V}$,
Eqs.~(\ref{eq:q_moment_1})-(\ref{eq:q_moment_3}) demonstrate that the first
three moments of the potential are also diagonal under the IG assumption of
Eq.~(\ref{eq:mat_S_delta}). More generally, the text following
Eq.~(\ref{eq:v_ab_2}) shows that each term of the Taylor expansion of
$V_{\alpha\beta}$ contains a~factor of $S_{\alpha\beta}$; hence if
$S_{\alpha\beta}=\delta_{\alpha\beta}$ then $V_{\alpha\beta}\propto
\delta_{\alpha\beta}$ for any well-behaved potential $V(q)$. In practice,
however, Eq.~(\ref{eq:mat_S_delta}) is satisfied only approximately and higher
order terms in $V$ may lead to significant couplings even when the overlaps
are small. Approximate diagonality of $\mathbf{H}$ must therefore be taken as
an additional assumption, which we consider to be an inherent part of the IG approximation.

Employing the IG approximation in Eq.~(\ref{eq:working_eq}) thus switches off
communication between basis functions as in the Independent Gaussian
Approximation.\cite{Sawada_85} More explicitly, since elements of $\mathbf{D}$
satisfy
\begin{equation}
D_{\alpha\beta}(t)=-\frac{i}{\hbar}{p_{\alpha}^{t}}^{\mathrm{T}}\!\cdot
{\dot{q}_{\alpha}^{t}}\,\delta_{\alpha\beta}, \label{eq:mat_D_delta}%
\end{equation}
propagation equation~(\ref{eq:working_eq}) decouples,
\begin{align}
\dot{c}_{\alpha}(t)  &  =\frac{i}{\hbar}\left[  i\hbar\,D_{\alpha\alpha
}(t)-H_{\alpha\alpha}(t)\right]  c_{\alpha}(t)\nonumber\\
&  =\frac{i}{\hbar}\left[  {p_{\alpha}^{t}}^{\mathrm{T}}\!\cdot{\dot
{q}_{\alpha}^{t}}-H_{\alpha\alpha}(t)\right]  c_{\alpha}(t),
\end{align}
and one can formally write the solution as
\begin{equation}
c_{\alpha}(t)=c_{\alpha}\exp\left\{  \frac{i}{\hbar}\int_{0}^{t}\!d{\tilde{t}%
}\left[  {p_{\alpha}^{\tilde{t}}}^{\mathrm{T}}\!\cdot{\dot{q}_{\alpha}%
^{\tilde{t}}}-{H}_{\alpha\alpha}(\tilde{t})\right]  \right\}  ,
\label{eq:coeff_c_delta}%
\end{equation}
where the kinetic contribution [Eq.~(\ref{app:eq:mat_T})] to ${H}%
_{\alpha\alpha}(t)$ is%
\[
T_{\alpha\alpha}(t)=\frac{1}{2}\,{p_{\alpha}^{t}}^{\mathrm{T}}\!\cdot{m^{-1}%
}\cdot{p_{\alpha}^{t}+}\frac{\hbar^{2}}{4}{\mathrm{Tr}}(\gamma\,\cdot m^{-1})
\]
and $m$ denotes the diagonal mass matrix. This result is equivalent to the
central equation of Heller's FGA.\cite{Heller_81}

\textit{Approximating }$V_{\alpha\beta}$. In \textit{ab~initio\/}
applications, the most challenging part of the propagation
(\ref{eq:working_eq}) is evaluating potential matrix elements $V_{\alpha\beta
}$. Even for analytical potentials, it is generally impossible to obtain
$V_{\alpha\beta}$ in closed form. Thanks to the local nature of Gaussian basis
states~(\ref{eq:gaussian}), a~useful approximation is provided by expanding
the potential in a~Taylor series and evaluating the resulting integrals
analytically. Most frequently used are the following two approximations:

1.~\textit{Local Approximation} (LA). Taylor expansion~(\ref{eq:v_ab_2}) is
truncated after the zeroth order:
\begin{equation}
V_{\alpha\beta}\approx V_{\alpha\beta}^{\text{LA}}=V(Q)S_{\alpha\beta},
\label{eq:local_app}%
\end{equation}
where $Q=(q_{\alpha}+q_{\beta})/2$. The LA is equivalent to the zeroth-order
saddle-point approximation used in Multiple Spawning.\cite{Martinez_ACP_02}

2.~\textit{Local Harmonic Approximation} (LHA). Taylor
expansion~(\ref{eq:v_ab_2}) is truncated after the second order, using
Eqs.~(\ref{eq:q_moment_1}) and~(\ref{eq:q_moment_2}) for the first and second
moments. The diagonal elements, in particular, become%
\begin{equation}
V_{\alpha\alpha}^{\text{LHA}}=V(q_{\alpha})+\frac{1}{4}{\mathrm{Tr}}\left[
\nabla^{2}V(q_{\alpha})\cdot\,\gamma^{-1}\right]  . \label{eq:mat_TV_delta}%
\end{equation}
As a~result, the LHA requires the expensive Hessian matrix.
\citeauthor{Sawada_85} observed that the LHA in conjunction with the
variational principle decouples the Gaussian basis functions. In our setting,
this decoupling effect\cite{Sawada_85,Burghardt_99} does not occur both
because the Gaussians are not treated fully variationally (in particular, we
use frozen and not thawed Gaussians) and because the potential is expanded
about the average coordinate $Q=(q_{\alpha}+q_{\beta})/2$ instead of
$q_{\alpha}$ or $q_{\beta}$ [see Eq.~(\ref{eq:v_ab_2})].
Therefore we distinguish between the Independent Gaussian Approximation of
\citeauthor{Sawada_85} and IG of Eq.~(\ref{eq:mat_S_delta}).

Our numerical calculations also exploited the fourth-order expansion,
permitting exact treatment of the potential in all systems discussed in
Sec.~\ref{sec:results}.

\textit{Evolving the basis with the average Hamiltonian }(AH). Unlike IG or
LHA, this approximation is specific to the fidelity amplitude. In analogy to
the DR, a~single basis was employed for the two propagations and evolved
classically with the \textit{average Hamiltonian }$H:=(H_{\text{I}%
}+H_{\text{II}})/2$ according to Eq.~(\ref{eq:classical}). With this
assumption $\mathbf{M}\equiv\mathbf{S}$ and fidelity
amplitude~(\ref{eq:fid_12}) simplifies to
\begin{equation}
f_{\text{AH}}(t,\,\tau)=\mathbf{c}_{1}(t,\,\tau)^{\dagger}\,\mathbf{S}%
(t,\,\tau)\,\mathbf{c}_{0}(t,\,\tau)=:f_{\text{GDR}}(t,\tau),
\label{eq:fid_12_avg}%
\end{equation}
where the subscripts on $\mathbf{c}$'s must be retained since the Hamiltonian
matrix elements in Eq.~(\ref{eq:working_eq}) still depend on the electronic
state. Note that if the basis were complete at all times, using the AH would
not constitute any approximation. There is an important difference, however,
between the GBM~(\ref{eq:fid_12}) and AH method~(\ref{eq:fid_12_avg}). In GBM,
$|f|$ decays partially due to decreasing overlap between corresponding basis
functions, i.e., due to decreasing diagonal elements of $\mathbf{M}$. In the
AH method~(\ref{eq:fid_12_avg}), in contrast, a~single basis is used, and the
diagonal elements of the overlap matrix $\mathbf{S}$ remain unity at all
times. Hence $|f|$ decays exclusively due to interference and not due to basis
overlaps. A~similar interpretation gave the name to the semiclassical
Dephasing Representation (\ref{eq:fidelity_amp_DR}), in which $|f|$ decays
solely due to dephasing and not due to decreasing classical
overlaps.\cite{Vanicek_04,Vanicek_06} Because of this analogy, we refer to the
AH\ method specified by Eq.~(\ref{eq:fid_12_avg}) as the \textit{Gaussian
Dephasing Representation} (GDR). Note that the idea of using a common basis
was also exploited in the \textquotedblleft single-set\textquotedblright%
\ version of the MCTDH method~\cite{book_MCTDH,*Meyer:1990} and in
Shalashilin's multiconfigurational Ehrenfest method,\cite{Shalashilin:2010}
where the common basis is propagated with a Hamiltonian given by an
Ehrenfest-weighted average instead of an arithmetic average of $H_{\text{I}}$
and $H_{\text{II}}$ as in the GDR.

\begin{figure}
[htbp]\includegraphics[width=3.25in]{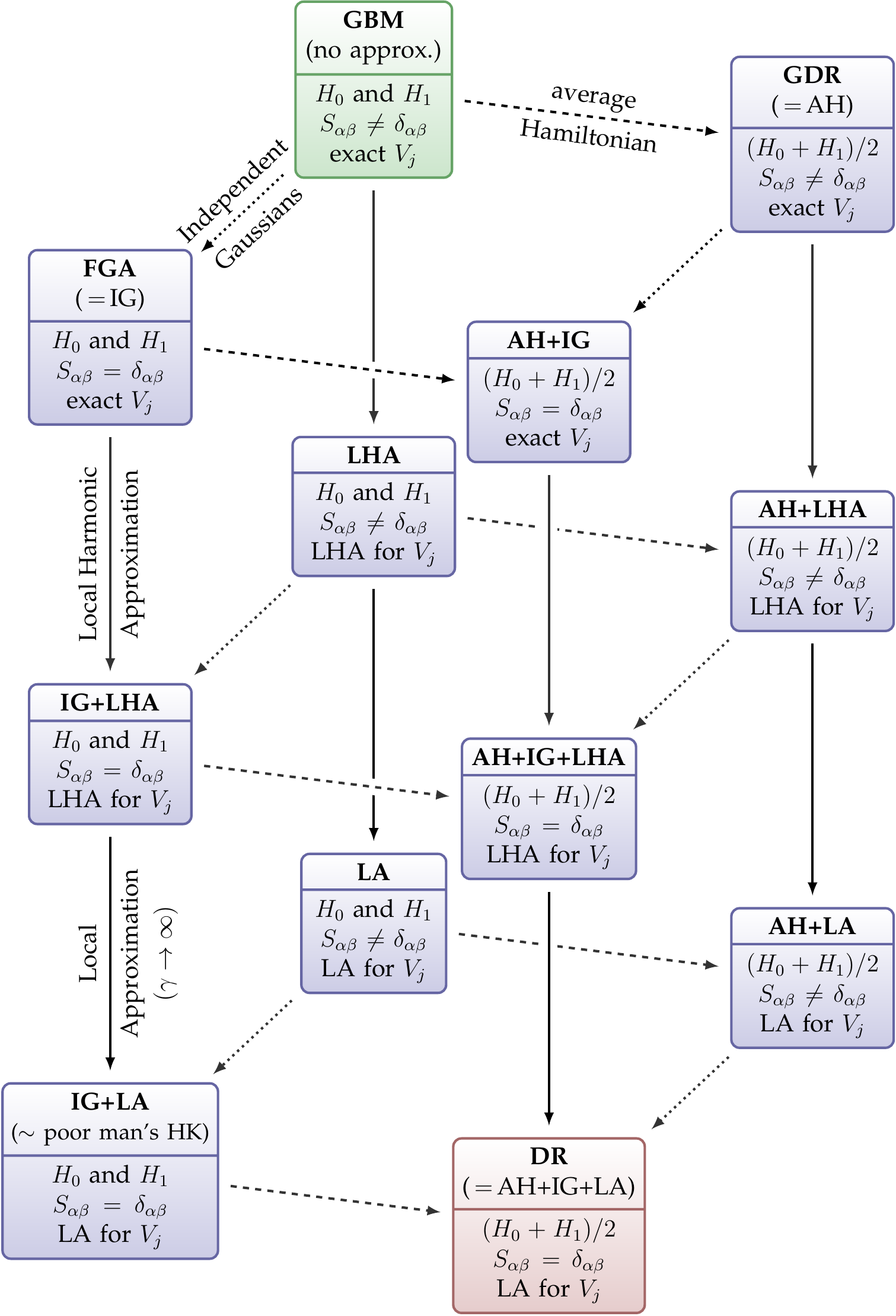} \caption{\label{fig:scheme_3d}
Approximations involved in the derivation of Dephasing Representation (DR) from the Gaussian Basis Method
(GBM). The commutative diagram shows several ways to go from the--in principle exact--GBM (top) to the--usually
least accurate but most efficient--DR (bottom). Different
approximations are distinguished by different directions and different line types: dotted lines = Independent Gaussians
(IG), dashed lines = average Hamiltonian (AH), and solid lines
= Local Harmonic Approximation (LHA) or Local Approximation (LA) for the
potential energy matrix elements. The figure indicates that the renormalized IG+LA value agrees
in absolute value with poor man's Herman-Kluk approximation.\cite{Tatchen_11}}
\end{figure}

\textit{Derivation of the DR from GBM. }We now derive the DR from the exact
GBM ~(\ref{eq:fid_12}) by a~sequence of four approximations: AH, IG, LHA, and
LA (see Fig.~\ref{fig:scheme_3d}). As shown above, AH approximation yields the
GDR~(\ref{eq:fid_12_avg}), which, together with the IG
approximation~(\ref{eq:mat_S_delta}), gives
\begin{align}
f_{\text{AH+IG}}(t,\,\tau)  &  =\mathbf{c}_{1}(t,\,\tau)^{\dagger}%
\,\mathbf{c}_{0}(t,\,\tau)\nonumber\\
&  =\sum_{\alpha}|c_{\alpha}|^{2}\exp\left[  \frac{i}{\hbar}\int_{\tau
}^{t+\tau}\!\!\!\!\!\!\!\!\!d{\tilde{t}}\,\Delta V_{\alpha\alpha}(q_{\alpha
}^{{\tilde{t}}})\right]  . \label{eq:fid_12_avg_iga}%
\end{align}
The LHA (\ref{eq:mat_TV_delta}) implies that $\Delta V_{\alpha\alpha}%
\approx\Delta V_{\alpha\alpha}^{\text{LHA}}:=V_{0,\alpha\alpha}^{\text{LHA}%
}-V_{1,\alpha\alpha}^{\text{LHA}}$ and%
\begin{equation}
f_{\text{AH+IG+LHA}}(t,\tau)=\sum_{\alpha}|c_{\alpha}|^{2}\exp\left[  \frac
{i}{\hbar}\int_{\tau}^{t+\tau}\!\!\!\!\!\!\!\!\!d{\tilde{t}}\,\Delta
V_{\alpha\alpha}^{\text{LHA}}(q_{\alpha}^{{\tilde{t}}})\right]  .
\label{eq:f_delta}%
\end{equation}
Finally, using the cruder LA (\ref{eq:local_app}) instead of the LHA implies
that $V_{j,\alpha\alpha}(q_{\alpha}^{t})\approx V_{j}(q_{\alpha}^{t})$ and
\begin{align}
f_{\text{AH+IG+LA}}(t,\,\tau)  &  =\sum_{\alpha}\,\lvert{c}_{\alpha}\rvert
^{2}\exp\left[  \frac{i}{\hbar}\int_{\tau}^{t+\tau}\!\!\!\!\!d{\tilde{t}%
}\,\Delta V(q_{\alpha}^{\tilde{t}})\right] \nonumber\\
&  =f_{\text{DR}}(t,\,\tau), \label{eq:f_delta_2}%
\end{align}
which is a~discretized version of the DR~(\ref{eq:fidelity_amp_DR}) with the
square coefficients $\lvert{c}_{\alpha}\rvert^{2}$ playing the role of the
Monte Carlo sampling weights. Note that the last result could also be obtained
by assuming infinitesimally narrow Gaussians,\ i.e., $\gamma\rightarrow\infty
$, in Eq.~(\ref{eq:mat_TV_delta}).

Derivation of the DR from the GBM is summarized in Fig.~\ref{fig:scheme_3d},
showing the four elementary approximations involved. Since three
approximations may be taken in arbitrary order, ten intermediate methods exist
between the GBM and DR, all together giving twelve methods ranging from the
exact GBM to the semiclassical DR. Above, final expressions were presented for
six of the twelve methods: GBM~(\ref{eq:fid_12}), FGA = IG
[Eqs.~(\ref{eq:fid_12}) and (\ref{eq:coeff_c_delta})], GDR\thinspace
=\thinspace AH~(\ref{eq:fid_12_avg}), AH+IG~(\ref{eq:fid_12_avg_iga}),
AH+IG+LHA~(\ref{eq:f_delta}), and DR = AH+IG+LA~(\ref{eq:f_delta_2}). The
remaining six methods are easily obtained by applying a~subset of the four
elementary steps (AH, IG, LHA, or LA) to the original GBM~(\ref{eq:fid_12}).

\section{Numerical examples\label{sec:results}}

In this section, the methods from Sec.~\ref{sec:theory} are used to compute
time correlation functions required for TRSE spectra. While the efficiency of
the original DR is pronounced especially in high-dimensional systems (due to
its $D$-independent convergence rate mentioned in the Introduction), here we
focus on few-dimensional systems which permit benchmark exact quantum
calculations using the \textit{Thawed Gaussian Approximation}%
\cite{Heller_75,Lee_82} (TGA) or split-operator methods. As pointed out above,
the GDR and GBM constitute a~conceptual bridge between computational
efficiency and formal accuracy; while both GDR and GBM converge to the exact
quantum result, the number of trajectories required for convergence will
certainly increase with $D$, yet---as will be clear from the examples
below---this growth is typically much slower than the exponential scaling with
$D$ for fixed-grid methods.


\subsection{Test systems\label{subsec:test_systems}}


\textit{Harmonic potential}. In this model, the potential energy surface $j$
is represented by a~one-dimensional harmonic potential
\begin{equation}
V_{j}(q)=E_{j}+\frac{1}{2}k_{j}(q-d_{j})^{2}, \label{eq:pot_harmonic}%
\end{equation}
where $d_{j}$ is the displacement and $k_{j}$ the force constant. (For
convenience, we set $E_{0}=E_{1}=0$, since nonzero $E_{j}$ values only shift
the spectrum, but do not change its shape.)

\textit{Pyrazine $S_{0}/S_{1}$ model. }This system is a~simplified version of
the four-dimensional vibronic coupling model, which takes into account normal
modes $\nu_{1}$, $\nu_{6a}$, $\nu_{9a}$, and $\nu_{10a}$ of
pyrazine.\cite{Stock_95} The $S_{0}$ and $S_{1}$ surfaces from
Ref.~\onlinecite{Stock_95} are used, but the nonadiabatic coupling between
states $S_{1}$ and $S_{2}$ is neglected since this coupling is much less
important for the $S_{0}\rightarrow S_{1}$ excitation than for the often
studied $S_{0}\rightarrow S_{2}$ excitation. This approximation was justified
by two independent exact quantum calculations, with and without the
$S_{1}/S_{2}$ coupling, which yielded only marginally different spectra (not
shown).
However, even this simplified model requires a~nontrivial Duschinsky
rotation\cite{Duschinsky_1937} to transform between normal modes of the ground
and excited states.

\textit{Quartic oscillator.} This two-dimensional system, chosen because of
its chaotic
dynamics,\cite{Meyer_86,*Waterland_88,*Pollak_89,*Bohigas_93,*Revuelta_12}
consists of two potential energy surfaces
\begin{equation}
V_{j}(q_{1},q_{2})=E_{j}+\frac{1}{2}q_{1}^{2}q_{2}^{2}+\frac{1}{4}\beta
_{j}(q_{1}^{4}+q_{2}^{4}). \label{eq:pot_quartic}%
\end{equation}
Chaotic behavior is due to the coupling term $q_{1}^{2}q_{2}^{2}/2$ since for
$\beta_{j}\rightarrow\infty$ the Hamiltonian $T+$ $V_{j}$ becomes separable
and hence integrable.

\subsection{Computational details}

The initial state was a Gaussian representing the ground vibrational state of
the ground PES $V_{0}$ [in the harmonic potential (\ref{eq:pot_harmonic})] or
$S_{0}$ (in the pyrazine $S_{0}/S_{1}$ model). Initial states used in the
quartic oscillator are specified in the figure captions. The Gaussian basis
was generated with the Monte Carlo technique (with $\xi=2$ and $\epsilon=0.8$,
see Appendix~\ref{subsec:basis_choice}) except in one-dimensional
applications, where an equidistant phase-space grid was used (with $N_{q}=$
$N_{p}=N^{1/2}$, see Appendix~\ref{subsec:basis_choice}). The width matrix
$\gamma$ from Eq.~(\ref{eq:basis_fn}) was always equal to the width matrix
$\Gamma$ of the initial state.

In the quartic oscillator, exact quantum-mechanical (QM)\ benchmark results
were obtained with a~fourth-order split-operator method,\cite{Wehrle_11}
whereas in the harmonic and pyrazine $S_{0}/S_{1}$ models, QM results were
obtained with Heller's TGA (Refs.~\onlinecite{Heller_75,Lee_82}), which is exact in
quadratic potentials. Classical trajectories were evolved with a~fourth-order
symplectic integrator,\cite{Wehrle_11} while the
propagation~(\ref{eq:working_eq_d}) of the $\mathbf{c}$ vector was performed
with the exponential method~(\ref{eq:method_exp}). This was feasible because
of the limited size of the basis. The propagation time steps used for the
harmonic oscillator, pyrazine, and quartic oscillator were $1$~a.u.,
$0.2$~a.u., and $10^{-3}$, respectively. Unit mass was used unless stated otherwise.

For Gaussian initial states in quadratic potentials, the DR results were
computed efficiently with the recently published\cite{Sulc_MolPhys_12}
\textit{Cellular Dephasing Representation} (CDR). This was possible since
under these conditions CDR based on a~single trajectory is equivalent to the
fully converged DR, which would otherwise require thousands of
trajectories.\cite{Zambrano_tba}


\subsection{\label{subsec:num_results}Time-resolved stimulated emission: time
correlation functions and spectra}

Now we turn to the main results comparing the approximations discussed in
Subsec.~\ref{subsubsec:approx}. Three overall conclusions can be drawn from
these results:\ First, the GBM corrects the inaccuracies of the DR. Second, in
chaotic systems, a~finite basis evolving with the average Hamiltonian can,
surprisingly, provide more accurate results than two bases evolved separately.
Third, despite its simplicity, even the original DR is useful for computing
TRSE spectra. The results are presented in four groups according to whether
the DR works and whether the GBM~(\ref{eq:fid_12}) converges faster than the
GDR~(\ref{eq:fid_12_avg}).

\begin{figure*}[ptbh]%
\includegraphics[width=\hsize]{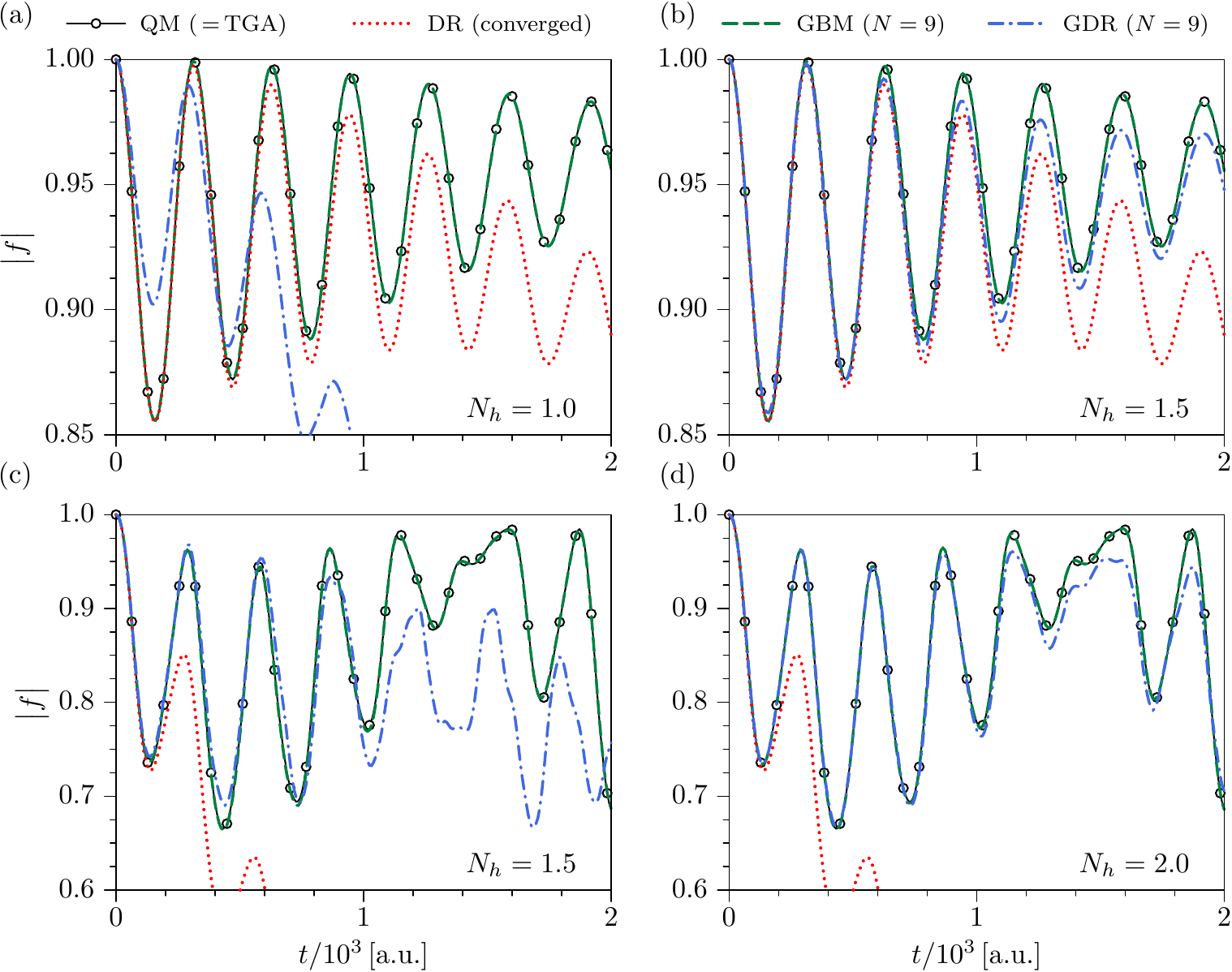}\caption{Time correlation function for
time-resolved stimulated emission between two harmonic
surfaces~(\ref{eq:pot_harmonic}). Delay time $\tau=5000\,\text{a.u.}\approx121\,\text{fs}$, mass $m=1250\,$a.u., and the initial state is the
ground state of the ground surface $V_{0}$. $N_{h}$ gives the number of basis
functions per phase-space area $h=2\pi\hbar$. Parameters in
Eq.~(\ref{eq:pot_harmonic}) are (all in a.u.): $E_{0}=E_{1}=d_{0}=0$,
$d_{1}=0.08$, $k_{0}=0.5$, and (a/b)~$k_{1}=0.52$: displacement is the
dominant change; (c/d)~$k_{1}=0.7$: change of the force constant is dominant.}\label{fig:lho_dq}%
\end{figure*}

\textit{GBM outperforms GDR and DR works. }This occurs, e.g., in the harmonic
model~(\ref{eq:pot_harmonic}) with a~large displacement $\left\vert
d_{1}-d_{0}\right\vert $ and only a~small change in the force constant
$\left\vert k_{1}-k_{0}\right\vert $ [see Fig.~\ref{fig:lho_dq}(a)-(b)]. The
figure shows that both GBM and GDR converge to the exact QM result. As
expected in this simple system, GBM converges faster than GDR, in which the
basis evolves with the average Hamiltonian. Convergence of GDR is accelerated
by moving the $N$ Gaussians closer to one another [compare panels (a) and
(b)]. While our grid is regular, a similar effect was observed in
\textquotedblleft compressed\textquotedblright\ Monte Carlo
sampling.\cite{DVS_pancakes} Even the DR is rather accurate and would be
exact\cite{Mukamel_82} for $k_{1}=k_{0}$.


\textit{GBM outperforms GDR and DR breaks down. }The DR breaks down in simple
systems such as the harmonic surfaces~(\ref{eq:pot_harmonic}) when the force
constants differ significantly. Figure~\ref{fig:lho_dq}(c) shows that the DR
captures the initial decay of $f$ but not its revivals. Methods employing
Gaussian bases fix this failure. GBM converges faster than GDR, although the
performance of GDR is, again, improved if the basis functions are closer to
one another [compare panels (c) and (d)]. Another way to partially correct the
breakdown of DR is to multiply the contributions to the DR by
trajectory-dependent prefactors.\cite{Zambrano_PRE_11,Zambrano_tba}
Fortunately, in real systems with dissipation the recurrences in $f$ are
damped, which improves the credibility of the DR.

\begin{figure}
[htbp]\includegraphics[width=3.25in]{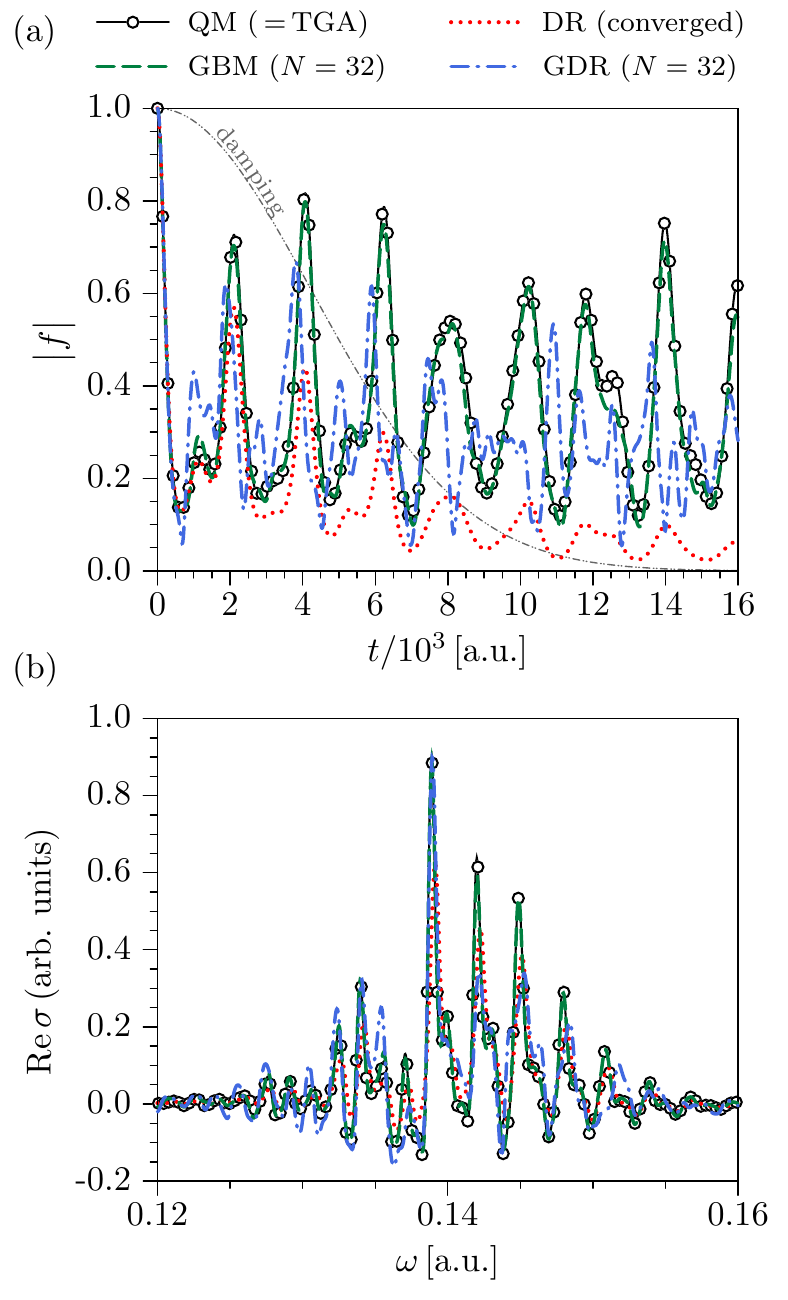} \caption{
\label{fig:pyrazine_delay}
Time-resolved stimulated emission in the pyrazine $S_{0}/S_{1}$  model from Subsec.~\ref{subsec:test_systems}. Initial state is
the ground state of the $S_0$ surface and the delay time $\tau=2\cdot10^3\,\text{a.u.}\approx48\,\text{fs}$.
(a)~Time correlation function. The damping function  of
Eq.~(\ref{eq:damping}) with $T=6\cdot10^3\,\text{a.u.}\approx145\,\text{fs}$ is shown by a~gray
dash-double-dotted line.
(b)~Corresponding spectrum.
}
\end{figure}

As shown in Fig.~\ref{fig:pyrazine_delay}(a), in the pyrazine $S_{0}/S_{1}$
model the GBM again converges to the exact QM\ result faster (with $N=32$)
than the GDR. Although the DR does not yield correct amplitudes of the peaks,
their positions are reproduced remarkably well. This is further confirmed in
the TRSE spectrum in Fig.~\ref{fig:pyrazine_delay}(b), which was computed by
Fourier transforming $f$ multiplied by a~phenomenological damping
function\cite{book_MCTDH,*Meyer:1990}
\begin{equation}
\chi(t)=\exp(-t^{2}/T^{2}). \label{eq:damping}%
\end{equation}
As the positions of peaks are well reproduced, one could consider this a
success rather than a~failure of DR. Note that the negative values present
even in the exact QM spectrum in Fig.~\ref{fig:pyrazine_delay}(b) are not
numerical artifacts---unlike a continuous-wave spectrum, the TRSE spectrum
defined by Eq.~(\ref{eq:sigma_spec}) is not guaranteed to be positive for all
frequencies, although its integral over all frequencies is positive. The
appearance of negative values is due to the nonstationary character of the
initial state prepared by the pump pulse on the excited surface, and is
discussed in detail in Ref.~\onlinecite{Pollard_Mathies:1990}.

\begin{figure}
[htbp]\includegraphics[width=3.25in]{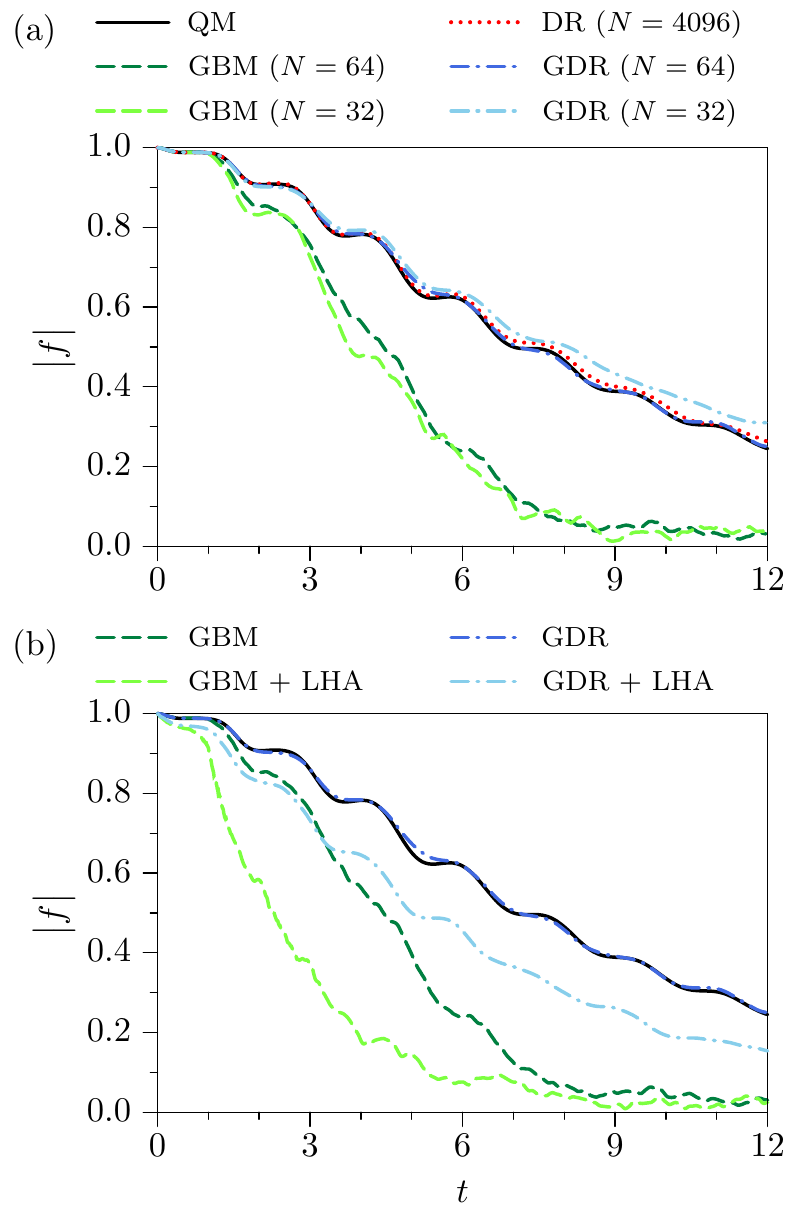}
\caption{\label{fig:quartic_dr_ok}Time correlation function for time-resolved stimulated emission in quartic
oscillator~(\ref{eq:pot_quartic}). Parameters of
the potential energy surfaces are $E_{0} = E_{1} = 0$ and,  in the notation of Eq.~(\ref{eq:delta}),
$\beta_0=0.2$ and
$\delta=1/16$.
Time delay $\tau = 0$. Initial state is a~Gaussian~(\ref{eq:basis_fn}) with $\gamma_{1}=\gamma_{2}=1$ centered at
$(Q_{\text{init}},\,P_{\text{init}})$, where $Q_{\text{init}}=(0,4)$ and $P_{\text{init}}=(4,0)$.
(a)~Comparison of various methods and their convergence as a~function of $N$: GDR converges faster than GBM.
($V$ is treated exactly in both methods.)
(b)~Effect of the Local Harmonic Approximation (LHA) on GBM and GDR:
The LHA for $V$ breaks down.
($N=64$ in all methods. In the notation of Fig.~\ref{fig:scheme_3d},
GBM+LHA stands for LHA and GDR+LHA for AH+LHA. )
}
\end{figure}

\textit{GDR outperforms GBM and DR works.} Unexpectedly, in the chaotic
quartic oscillator~(\ref{eq:pot_quartic}) the seemingly more approximate GDR
converges faster than the GBM [see Fig.~\ref{fig:quartic_dr_ok}(a)]. Although
the rapid divergence of classical trajectories aggravates the incompleteness
of the Gaussian basis, this problem is much less severe in GDR. In GBM, the
two bases diverge rapidly even for a~small change $\left\vert \beta_{1}%
-\beta_{0}\right\vert $ in Eq.~(\ref{eq:pot_quartic}). Unless both bases cover
essentially the entire available phase space, $f_{\text{GBM}}$ will decay
artificially fast due to the decay of overlaps between the two bases. In
contrast, GDR avoids this decay by using a~single basis for dynamics on both
surfaces. Unlike the GBM, which would only converge when the two bases
approached completeness, the GDR converges with a very small basis since the
main contribution to the decay of $f(t)$ in the chaotic quartic oscillator
comes from the decay of the scalar product $\mathbf{c}_{1}(t,\,\tau)^{\dagger
}\,\mathbf{c}_{0}(t,\,\tau)$ and hence is relatively insensitive to the
off-diagonal elements of the overlap matrix $\mathbf{S}$. Similarly, the
success of the original DR in chaotic systems relies on the use of a~single
Hamiltonian for propagating classical
trajectories.\cite{Vanicek_04,Vanicek_06} This explanation is confirmed in
Fig.~\ref{fig:quartic_dr_ok}(a), in which GBM exhibits a~spurious decay,
whereas GDR (with $N=64$) agrees with the quantum result. Remarkably, due to
chaotic motion, increasing $N$ up to $256$ improves the GBM\ result only
marginally (not shown). Thanks to using the average Hamiltonian, DR matches
the quantum dependence as well as the GDR, albeit at the cost of more
trajectories. Comparing GDR with and without the LHA for the potential,
Fig.~\ref{fig:quartic_dr_ok}(b) demonstrates that the widely used LHA breaks
down completely in the quartic oscillator. In this system, matrix elements of
$V$ must be treated exactly.

\begin{figure*}
[ptbh]\includegraphics[width=\hsize]{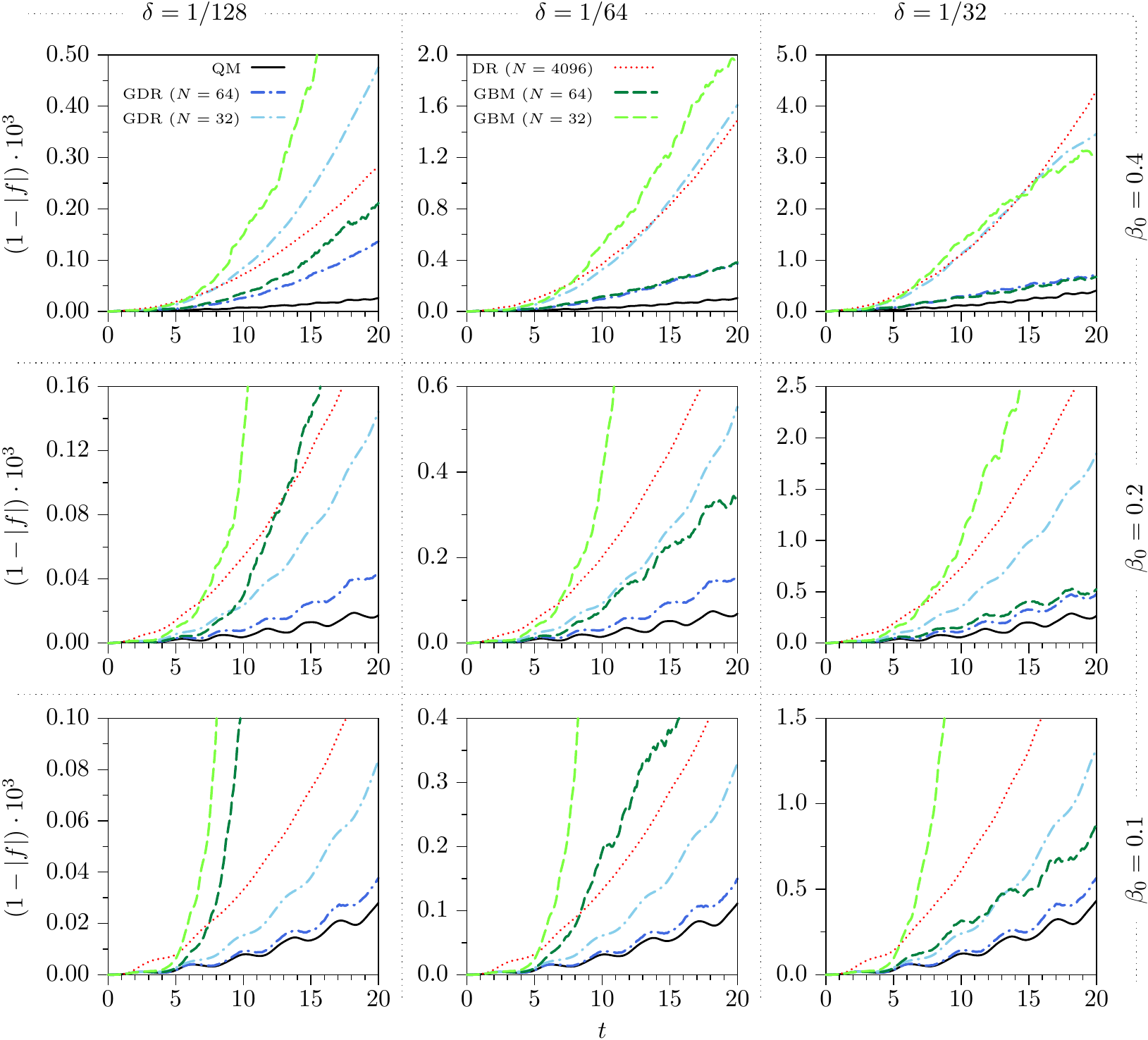}
\caption{Time correlation function for
time-resolved stimulated emission in quartic oscillator~(\ref{eq:pot_quartic}). Time delay $\tau=0$. Initial state is a~Gaussian~(\ref{eq:basis_fn}) with
$\gamma_{1}=\gamma_{2}=1$ centered at the phase-space origin. Parameter
$\delta$ is the relative perturbation strength defined by Eq.~(\ref{eq:delta}), whereas $\beta_{0}$ controls chaoticity
(which increases with decreasing $\beta_{0}$). GBM and GDR results are averages over 10 realizations.}\label{fig:quartic_dr_ko}%
\end{figure*}

\textit{GDR outperforms GBM and DR breaks down.} The success of the DR in
chaotic systems is not universal. Figure~\ref{fig:quartic_dr_ko} shows the
time correlation function for several choices of the parameter $\beta_{0}$,
controlling chaoticity, and of the difference $\Delta\beta:=\beta_{1}%
-\beta_{0}$ between the two surfaces. In Fig.~\ref{fig:quartic_dr_ko}, the
perturbation strength is measured by parameter $\delta$, defined as
\begin{equation}
\delta:=\Delta\beta/\beta_{0}=\beta_{1}/\beta_{0}-1. \label{eq:delta}%
\end{equation}
In eight of the nine Fig.~\ref{fig:quartic_dr_ko} panels, GDR converges faster
than GBM. In accordance with the heuristic arguments presented above,
superiority of GDR over GBM increases with increasing chaoticity, i.e.,
decreasing $\beta_{0}$. This superiority disappears gradually with increasing
perturbation $\delta$ due to the increasing error inherent in the propagation
of the finite basis with the AH.

\section{\label{sec:conclusions}Conclusions}

We have shown how the DR~(\ref{eq:fidelity_amp_DR}), an efficient
semiclassical method for computing ultrafast electronic spectra, emerges
naturally from the formulation of quantum dynamics in a~classically evolving
Gaussian basis. This was achieved by a~series of three elementary
approximations: evolving the basis with the AH, using IG, and applying LA for
the potential. Along with the derivation based on linearizing the path
integral,\cite{Shi_05} this result puts the DR on strong theoretical footing
and justifies its place among efficient semiclassical methods for computing
specific time correlation functions. Moreover, the accuracy of the DR has been
increased by presenting two in-principle exact generalizations of the DR:\ the
GBM and GDR. The GBM is a~straightforward application of the concept of
a~Gaussian basis to time correlation functions of time-resolved spectroscopy.
The GDR, in contrast, is a~natural generalization of the DR since (i) GDR
utilizes a~single basis for propagating the quantum state with both
Hamiltonians, (ii) this basis propagates classically with the average
Hamiltonian, and (iii)\ the decay of the time correlation function is due to
interference and not due to decay of basis overlaps.

As expected, in many situations the GBM converges faster than the GDR.
Surprisingly, in chaotic systems the GDR can outperform the GBM in which the
two bases evolve separately with the \textquotedblleft
correct\textquotedblright\ Hamiltonians. Numerical results presented in
Sec.~\ref{sec:results} confirm that both methods achieve our main goal of
increasing the accuracy of the DR in calculations of ultrafast electronic
spectra. As a~by-product, ten intermediate methods between the GBM and DR have
been obtained, which may be useful for future applications. Relationships
between all twelve methods are shown in Fig.~\ref{fig:scheme_3d}, which also
represents the ubiquitous balancing between formal exactness (achieved
typically at a~high computational cost) on one hand and computational
efficiency on the other. In summary, we believe that our results provide
additional insight into the connections between various exact and
semiclassical methods, and demonstrate the practical value of semiclassical
and Gaussian basis approaches based on classical trajectories.
\begin{acknowledgments}
This research was supported by the Swiss NSF within the NCCR Molecular
Ultrafast Science and Technology (MUST) and by the EPFL. The authors would
like to thank T.~Zimmermann for providing the nonadiabatic spectrum of pyrazine.
\end{acknowledgments}
\appendix
\section{Gaussian integrals\label{app:A}}
Here we derive formulae for the $\mathbf{S}$, $\mathbf{D}$, and $\mathbf{H}$
matrix elements required in Eq.~(\ref{eq:working_eq}). All of these matrix
elements can be expressed in terms of three basic integrals
\begin{align*}
I_{0}(A,b)  &  :=\int\!d^{D}\!{q}\,\,e^{-{q}^{\mathrm{T}}\!\cdot{A}\cdot
{q}+{b}^{\mathrm{T}}\!\cdot{q}},\\
I_{1}(A,b,\delta)  &  :=\int\!d^{D}\!{q}\,\,({\delta}^{\mathrm{T}}\!\cdot
{q})\,e^{-{q}^{\mathrm{T}}\!\cdot{A}\cdot{q}+{b}^{\mathrm{T}}\!\cdot{q}},\\
I_{2}(A,b,\kappa)  &  :=\int\!d^{D}\!{q}\,\,({q}^{\mathrm{T}}\!\cdot{\kappa
}\cdot{q})\,e^{-{q}^{\mathrm{T}}\!\cdot{A}\cdot{q}+{b}^{\mathrm{T}}\!\cdot{q}%
},
\end{align*}
where $A$ and $\kappa$ denote $D\!\times\!D$ positive definite symmetric
complex matrices, while $b$, $c$, and $\delta$ are $D$-dimensional complex
vectors. The integral $I_{0}(A,b)$ is well known:\cite{book_ryzhik}%
\begin{equation}
I_{0}(A,b)=\left(  \frac{\pi^{D}}{\det A}\right)  ^{\frac{1}{2}}\exp\left(
\frac{1}{4}\,b^{\mathrm{T}}\!\cdot{A^{-1}}\cdot{b}\right)  . \label{eq:int_1}%
\end{equation}
Since $I_{1}(A,b,\delta)=\sum_{l=1}^{D}\delta_{l}\frac{\partial}{\partial
{}{b_{l}}}I_{0}(A,b)$, differentiation of Eq.\thinspace(\ref{eq:int_1}) with
respect to the components of $b$ gives
\begin{equation}
I_{1}(A,b,\delta)=\frac{1}{2}I_{0}(A,b)\left(  {\delta}^{\mathrm{T}}%
\!\cdot{A^{-1}}\cdot{b}\right)  . \label{eq:int_2}%
\end{equation}
Similarly one obtains
\begin{equation}
I_{2}(A,b,\kappa)=\frac{1}{4}I_{0}(A,b)\left[  c^{\mathrm{T}}\!\cdot{\kappa
}\cdot{c}+2\,\mathrm{Tr}\left(  \kappa\,\cdot A^{-1}\right)  \right]  ,
\label{eq:int_3}%
\end{equation}
where $c:={A^{-1}}\cdot{b}$, by noting that
\[
I_{2}(A,b,\kappa)=\sum_{l,k=1}^{D}\frac{\partial}{\partial{}{b_{l}}}%
\kappa_{lk}\frac{\partial}{\partial{}{b_{k}}}I_{0}(A,b).
\]

As mentioned in Sec.~\ref{sec:theory}, the Gaussian basis functions labeled by
index $\alpha$ have the form%
\begin{align}
\phi_{\alpha}(q)  &  = N_{\alpha}\exp\left[  -(q-q_{\alpha})^{\mathrm{T}%
}\!\cdot{\gamma}\cdot(q-q_{\alpha})/2\right. \nonumber\\
&  + \left.  i{p_{\alpha}}^{\mathrm{T}}\!\cdot(q-q_{\alpha})/\hbar\right]  ,
\label{app:eq:phi}%
\end{align}
where the superscript $t$ denoting time dependence is omitted for simplicity.
The $D\!\times\!D$ constant real matrix $\gamma$ is assumed to be independent
of $\alpha$ and to be symmetric positive definite in order that the basis
functions~(\ref{app:eq:phi}) be square normalizable.

Calculation of the overlap matrix $S_{\alpha\beta}=\langle\phi_{\alpha
}(t)|\phi_{\beta}(t)\rangle$ is simplified by introducing vectors
\begin{align*}
\Delta{}q  &  :=q_{\alpha}-q_{\beta}, & Q  &  :=(q_{\alpha}+q_{\beta})/2,\\
\Delta{}p  &  :=p_{\alpha}-p_{\beta}, & P  &  :=(p_{\alpha}+p_{\beta})/2.
\end{align*}
Special case of the integral $I_{0}$ [Eq.~(\ref{eq:int_1})] yields
\begin{align}
S_{\alpha\beta}  &  =\exp\left[  -\left(  \Delta{}q^{\mathrm{T}}\!\cdot
{\gamma}\cdot{\Delta{}q}+\Delta{}p^{\mathrm{T}}\!\cdot{\gamma^{-1}}\cdot
\Delta{}p/\hbar^{2}\right)  /4\right] \nonumber\\
&  \times\exp\left(  i\,{\Delta{}q}^{\mathrm{T}}\!\cdot{P/}\hbar\right)  .
\label{app:eq:mat_S}%
\end{align}
Application of the identity
\[
D_{\alpha\beta}=\left(  {\dot{q}_{\beta}}^{\mathrm{T}}\!\cdot{\nabla
_{\!q_{\beta}}}+{\dot{p}_{\beta}}^{\mathrm{T}}\!\cdot{\nabla_{\!p_{\beta}}%
}\right)  S_{\alpha\beta}%
\]
for the time-derivative matrix elements~(\ref{eq:mat_D_el}) to
Eq.~(\ref{app:eq:mat_S}) for $S_{\alpha\beta}$ gives
\begin{align}
D_{\alpha\beta}  &  =S_{\alpha\beta}\left[  {\dot{p}_{\beta}}^{\mathrm{T}%
}\!\cdot{\left(  \gamma^{-1}\cdot\Delta{}p+i\hbar\,{\Delta{}q}\right)
/(2\hbar^{2})}\right. \nonumber\\
&  \left.  +\,{\dot{q}_{\beta}}^{\mathrm{T}}\!\cdot{\left(  \gamma\cdot
{\Delta{}q/2}-iP/\hbar\right)  }\right]  . \label{app:eq:mat_D}%
\end{align}
As for the kinetic operator $\hat{T}$, we assume a~slightly generalized form
\begin{equation}
\hat{T}=-\frac{\hbar^{2}}{2}\sum_{k,l=1}^{D}g_{kl}\frac{\partial^{2}}%
{\partial{}q_{k}\partial{}q_{l}},
\end{equation}
where $g_{kl}$ represents matrix elements of a~symmetric positive definite
matrix $g$. (Nevertheless, all numerical calculations employed a~diagonal
$\hat{T}$ corresponding to $g_{kl}=\delta_{kl}m_{k}^{-1}$, i.e., $g$ was the
inverse of the diagonal mass matrix $m$.) Matrix elements of $\hat{T}$ are
given by
\begin{equation}
T_{\alpha\beta}=S_{\alpha\beta}\left[  \frac{1}{2}{P^{\text{eff}}}%
^{\mathrm{T}}\!\cdot{g}\cdot{P^{\text{eff}}}+\frac{1}{4}\hbar^{2}%
\mathrm{Tr}\left(  g\,\cdot\gamma\right)  \right]  \label{app:eq:mat_T}%
\end{equation}
with $P^{\text{eff}}:=P+i\hbar\,{\gamma}\cdot{\Delta{}q}/2$.

It is impossible to write the potential energy matrix elements $V_{\alpha
\beta}$ in a~closed form for a~general potential $V(q)$. One can, however,
obtain a~useful approximation by expanding the potential in a~truncated Taylor
series about the coordinate $Q$, at which the expression $\lvert\phi_{\alpha
}(q)^{\star}\phi_{\beta}(q)\rvert$ attains its maximum, and by evaluating the
resulting integral analytically. Adopting the multi-index notation, the
potential is approximated up to the $l$th order as
\begin{equation}
V(q)\approx\sum_{|\lambda|=0}^{l}\left.  \frac{D^{\lambda}V}{\lambda
!}\right\vert _{Q}\hspace*{-0.5em}\cdot(q-Q)^{\lambda}. \label{eq:v_ab_2}%
\end{equation}
Contributions to $V_{\alpha\beta}$ from individual terms in
Eq.~(\ref{eq:v_ab_2}) are obtained by a~repeated application of the
differential operator $i\hbar\left(  \nabla_{p_{\alpha}}-\nabla_{p_{\beta}%
}\right)  /2$ to the overlap matrix $S_{\alpha\beta}$. As a~result, one
obtains the first moment
\begin{align}
I_{\alpha\beta}  &  :=\langle\phi_{\alpha}\lvert\hat{q}-Q\rvert\phi_{\beta
}\rangle\nonumber\\
&  =\frac{i\hbar}{2}\left(  \nabla_{p_{\alpha}}-\nabla_{p_{\beta}}\right)
S_{\alpha\beta}=S_{\alpha\beta}\,\rho, \label{eq:q_moment_1}%
\end{align}
the second moment%
\begin{align}
J_{\alpha\beta,rs}  &  :=\langle\phi_{\alpha}\lvert(\hat{q}-Q)_{r}(\hat
{q}-Q)_{s}\rvert\phi_{\beta}\rangle\nonumber\\
&  =-\frac{\hbar^{2}}{4}\left(  \nabla_{p_{\alpha}}-\nabla_{p_{\beta}}\right)
_{rs}^{2}S_{\alpha\beta}\nonumber\\
&  =S_{\alpha\beta}\left[  \rho_{r}\rho_{s}+(1/2)\left(  \gamma^{-1}\right)
_{rs}\right]  , \label{eq:q_moment_2}%
\end{align}
and the third moment
\begin{align}
K_{\alpha\beta,rst}  &  :=\langle\phi_{\alpha}\lvert(\hat{q}-Q)_{r}(\hat
{q}-Q)_{s}(\hat{q}-Q)_{t}\rvert\phi_{\beta}\rangle\nonumber\\
&  =-\frac{i\hbar^{3}}{8}\left(  \nabla_{p_{\alpha}}-\nabla_{p_{\beta}%
}\right)  _{rst}^{3}S_{\alpha\beta}\nonumber\\
&  =S_{\alpha\beta}\{\rho_{r}\rho_{s}\rho_{t}+(1/2)[(\gamma^{-1})_{rs}\rho
_{t}\nonumber\\
&  \ \ +(\gamma^{-1})_{st}\rho_{r}+(\gamma^{-1})_{tr}\rho_{s}]\},
\label{eq:q_moment_3}%
\end{align}
where $\rho:=-i\gamma^{-1}\cdot{\Delta{}p/(}2\hbar)$. The fourth moment is
a~bit complicated to reproduce here. Nevertheless, it is easily evaluated by
using Eq.~(\ref{eq:q_moment_1}) together with
\[
\frac{i\hbar}{2}\left(  \nabla_{p_{\alpha}}-\nabla_{p_{\beta}}\right)
\rho=\frac{1}{2}\gamma^{-1}.
\]


\section{\label{sec:numerics}Efficient numerical implementation}


\subsection{\label{subsec:alg1}
\texorpdfstring{Numerical algorithm for the propagation equation~(\ref{eq:working_eq})}{Numerical algorithm for the propagation equation (18)}
}

\begin{algorithmic}[1]
\small
\Require{\Statex
\begin{itemize}[itemsep=0pt]
\item initial state $\lvert\Psi_{\text{init}}\rangle$ at time $t_0$, $\lvert\psi(t_0)\rangle=\lvert\Psi_{\text{init}}\rangle$
\item final propagation time $T$, time step $\Delta{}t$
\item number of basis elements $N$
\end{itemize}
}
\Ensure{state $\lvert\psi(T)\rangle$ at time $T$}
\State $t:=t_0$
\State expand $\lvert\Psi_{\text{init}}\rangle$ into the basis $\{\lvert\phi_\alpha(t)\rangle\}_{\alpha=1}^{N}$
\Comment{Eqs.~(\ref{eq:Psi_init})-(\ref{eq:psi})}
\While{$t\leq{}T$}
\State construct the $\mathbf{S}$,$\mathbf{D}$,$\mathbf{H}$ matrices
\Comment{Eq.~(\ref{eq:working_eq})}
\State update the expansion coefficients $c_\alpha(t)$
\State propagate classically all $N$ Gaussians by $\Delta{}t$
\State renormalize $\lvert\psi(t)\rangle$ for norm-nonconserving propagators
\State $t:=t+\Delta{}t$
\EndWhile
\end{algorithmic}

\subsection{\label{app:phase}
\texorpdfstring{Factoring out the semiclassical phase factor in Eq.~(\ref{eq:working_eq})}{Factoring out the semiclassical phase factor in Eq. (18)}
}

Propagation~(\ref{eq:working_eq}) can be accelerated by evaluating the
dominant oscillatory behavior of $c_{\alpha}(t)$ semiclassically, which is
achieved by factoring out the semiclassical phase factor in
expansion~(\ref{eq:psi}):%
\begin{equation}
|\psi(t)\rangle=\sum_{\alpha=1}^{N}\tilde{c}_{\alpha}(t)\,e^{iS_{\alpha
}^{\text{cl}}(t)/\hbar}\lvert\phi_{\alpha}(t)\rangle, \label{eq:psi_d}%
\end{equation}
where $S_{\alpha}^{\text{cl}}(t)=\int_{0}^{t}d\tilde{t}\,[\dot{q}_{\alpha
}^{\tilde{t}}\cdot p_{\alpha}^{\tilde{t}}-H(q_{\alpha}^{\tilde{t}},p_{\alpha
}^{\tilde{t}})]$ is the classical action. New coefficients $\tilde{c}_{\alpha
}(t)$ are propagated according to the equation
\begin{equation}
{\tilde{\mathbf{S}}(t)}\,{\dot{\tilde{\mathbf{c}}}(t)}=-{\left\{  \frac
{i}{\hbar}\left[  \tilde{\mathbf{H}}(t)+\bar{\mathbf{S}}(t)\right]
+\tilde{\mathbf{D}}(t)\right\}  }\,{\tilde{\mathbf{c}}(t)}.
\label{eq:working_eq_d}%
\end{equation}
The modified matrices can be expressed as
\begin{align}
{\tilde{Z}}_{\alpha\beta}(t)  &  =Z_{\alpha\beta}(t)\,\exp\left\{  \frac
{i}{\hbar}\left[  S_{\beta}^{\text{cl}}(t)-S_{\alpha}^{\text{cl}}(t)\right]
\right\}  \text{, }\\
{\bar{S}}_{\alpha\beta}(t)  &  =\tilde{S}_{\alpha\beta}(t)\dot{S}_{\beta
}^{\text{cl}}(t),
\end{align}
where $Z$ stands for $S$, $D$, or $H$. A~similar factorization was employed by
Mart\'{\i}nez and co-workers in the Multiple Spawning\cite{Martinez_ACP_02}
and by Shalashilin and co-workers in the Coupled Coherent
States.\cite{Shalashilin_JCP_00,*Shalashilin_JCP_01a,*Shalashilin_JCP_01b,*Shalashilin_ChemPhys_04}

\subsection{
\texorpdfstring{Solving the propagation Eqs.~(\ref{eq:working_eq}) or (\ref{eq:working_eq_d})}{Solving the propagation Eqs. (18) or (B2)}
}

Classical propagation of the basis (i.e., of $q_{\alpha}$ and $p_{\alpha}$)
and the action $S_{\alpha}^{\text{cl}}$ is performed with a~symplectic
algorithm.\cite{Brewer_97,Sulc_MolPhys_12} \ Quantum
propagation~(\ref{eq:working_eq}) of the basis coefficients $c_{\alpha}$ is
a~more involved process: While quite sophisticated
algorithms\cite{book_Koonin,book_NR,Levine:2008} exist for similar problems,
here we employed two simple methods based on dividing the propagation range
$[0,\,T]$ into equal intervals of size $\Delta{}t$.

Both the \textit{implicit Euler method,}
\begin{equation}
\mathbf{c}_{n+1}=\left[  \mathbf{S}_{n+1}+\left(  \frac{i}{\hbar}%
\mathbf{H}_{n+1}+\mathbf{D}_{n+1}\right)  \Delta{}t\right]  ^{-1}%
\!\!\mathbf{S}_{n+1}\mathbf{c}_{n}, \label{eq:method_eul}%
\end{equation}
and the \textit{\textquotedblleft exact\textquotedblright\ exponential
method,}%
\begin{equation}
\mathbf{c}_{n+1}=\exp\left[  -\mathbf{S}_{n}^{-1}\left(  \frac{i}{\hbar
}\mathbf{H}_{n}+\mathbf{D}_{n}\right)  \Delta{}t\right]  \mathbf{c}_{n},
\label{eq:method_exp}%
\end{equation}
are clear in the matrix notation, with subscript $n$ denoting the $n$th
propagation step. In addition, the wave function was renormalized by rescaling
$\mathbf{c}_{n}$ by $\left(  \mathbf{c}_{n}^{\dagger}\mathbf{S}_{n}%
\mathbf{c}_{n}\right)  ^{-1/2}$ after each step. While it was essential only
for the implicit Euler method, the renormalization was always performed since
for sufficiently large $N$, the $\mathcal{O}(N^{2}$) cost of renormalization
is negligible compared to the overall $\mathcal{O}(N^{3}$) cost of both
methods. In the algorithm of Subsec.~\ref{subsec:alg1}, steps 4, 5, and 6 must
be reordered as 6, 4, 5 for the implicit Euler method~(\ref{eq:method_eul})
since the basis must first be propagated in order to evaluate matrices at the
$(n+1)$th step.

In practice, it is neither necessary nor desirable to compute the inverse
matrix $\mathbf{S}_{n}^{-1}$. For example, rather than computing
$\mathbf{X:=S}_{n}^{-1}\left(  \frac{i}{\hbar}\mathbf{H}_{n}+\mathbf{D}%
_{n}\right)  $ as indicated, it is preferable to solve a~system of linear
equations $\mathbf{S}_{n}\mathbf{X}=\frac{i}{\hbar}\mathbf{H}_{n}%
+\mathbf{D}_{n}$ for $\mathbf{X}$ using any standard method. In this context,
Mart\'{\i}nez and co-workers suggested\cite{Martinez_JCP_96,Martinez_JPC_96}
to use singular value decomposition;\cite{book_NR} theoretical justification
was given\cite{Kay_89} by \citeauthor{Kay_89}. A~different approach,
consisting in inverting $\mathbf{S}_{n}$ by an iterative algebraic procedure
was explored\cite{Andersson_01} by \citeauthor{Andersson_01}.

Due to matrix exponentiation, the exponential method~(\ref{eq:method_exp}) is
feasible only for smaller basis sets. Although presumably exact, this method
does not necessarily permit a~larger time step $\Delta t$ than the first-order
implicit Euler method with renormalization. The reason is that for badly
conditioned $\mathbf{S}_{n}$, eigenvalues of the matrix $\mathbf{S}_{n}%
^{-1}\left(  \frac{i}{\hbar}\mathbf{H}_{n}+\mathbf{D}_{n}\right)  \Delta{}t$
become large except for very small time step $\Delta t$. Since most numerical
methods for matrix exponentiation require eigenvalues of the exponentiated
matrix to be small,\cite{EXPOKIT} lowering $\Delta t$ is required.

\citeauthor{Sawada_85} proposed\cite{Sawada_85} to monitor eigenvalues of
$\mathbf{S}_{n}$ during propagation, concluding that the basis size was
insufficient if all eigenvalues were (in absolute value) close to $1$. In
contrast, if some eigenvalues are very small, some functions should be removed
in order to restore regularity of $\mathbf{S}_{n}$. Such features, however,
have not been implemented here.

\begin{figure}
[htbp]\includegraphics[width=3.25in]{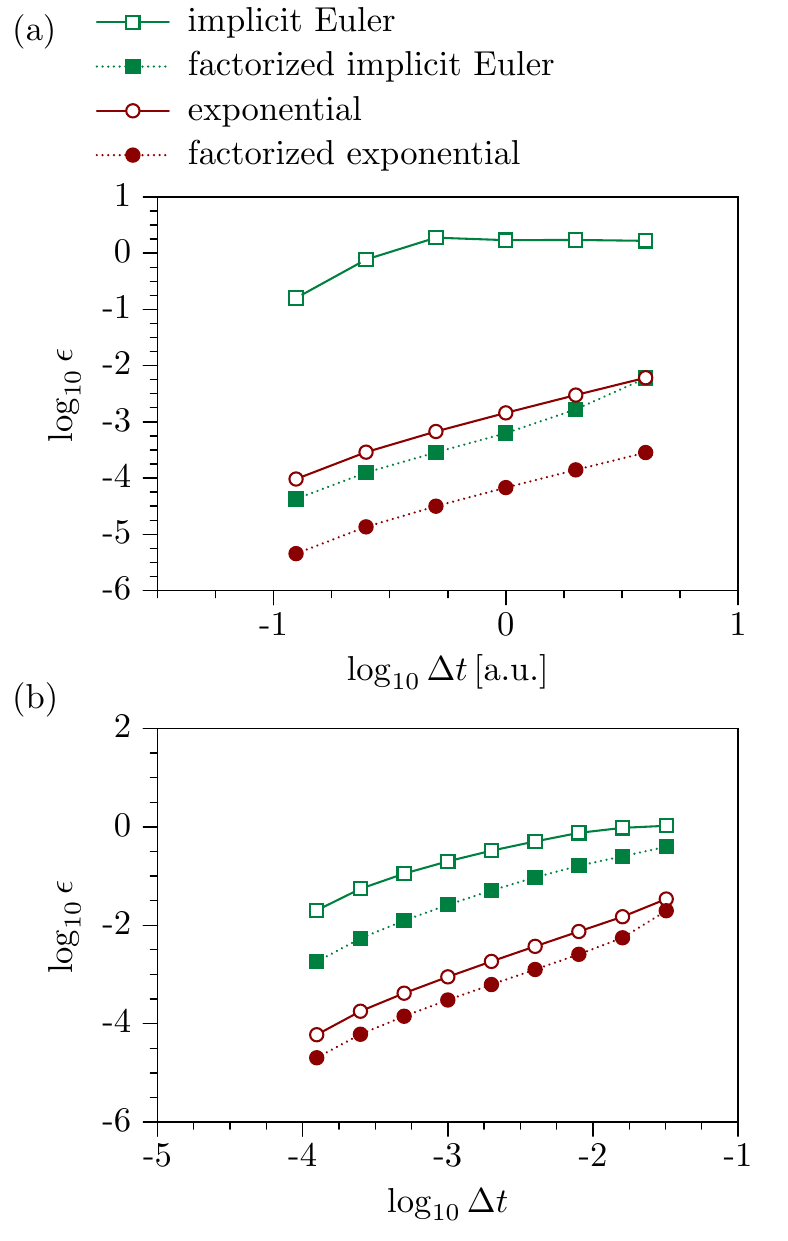} \caption{\label{fig:dt_conv}
Error of the correlation function $f$ computed with the GBM as a~function
of time step $\Delta t$. The figure compares the normalized $L^2$ errors
$\epsilon_{\Delta t}:=\left\Vert f_{\Delta t}-f\right\Vert /\left\Vert f\right\Vert$ on the time interval
$[0,T]$ obtained with the implicit Euler method (\ref{eq:method_eul}) and
exponential method (\ref{eq:method_exp})
applied to the propagation Eqs.~(\ref{eq:working_eq}) or (\ref{eq:working_eq_d})
[without and with the semiclassical factorization (\ref{eq:psi_d})].
(a)~Pyrazine $S_{0}/S_{1}$ model ($N=32$, $T=18000$~a.u., other parameters as in Fig.~\ref{fig:pyrazine_delay}).
(b)~Quartic oscillator ($N=32$, $T=12$, other parameters as in Fig.~\ref{fig:quartic_dr_ok}).
}
\end{figure}

In our calculations, the exponential method allows a~much larger time step
than the implicit Euler method (both in the pyrazine $S_{0}/S_{1}$ model and
in the quartic oscillator, see Fig.~\ref{fig:dt_conv}).
Figure~\ref{fig:dt_conv} also suggests that the modified propagation
Eq.~(\ref{eq:working_eq_d}) from Appendix~\ref{app:phase} permits increasing
the time step in comparison with the original propagation
Eq.~(\ref{eq:working_eq}). This improvement is more pronounced in the implicit
Euler method~(\ref{eq:method_eul}) than in the exponential
method~(\ref{eq:method_exp}).

\subsection{\label{subsec:basis_choice}
\texorpdfstring{Choice of the basis in Eq.~(\ref{eq:psi})}{Choice of the basis in Eq. (14)}
}
Another numerical issue is the choice of basis in Eq.~(\ref{eq:psi}), which
must represent $\lvert\Psi_{\text{init}}\rangle$ properly. A~small
approximation error (in the $L^{2}$ sense) in Eq.~(\ref{eq:psi}), however,
does not guarantee quality of the basis at later times. In general, a
compromise is required between the size of the basis and its acceptability
from the dynamical point of view.

Two methods used for the Sec.~\ref{sec:results} calculations are described
below, with more thorough discussions available
elsewhere.\cite{Burant_02,Davis_79,DVS_pancakes} To keep notation simple, we
assume that $D=1$ and consider the initial state~to be
a~Gaussian~(\ref{eq:basis_fn}):
\begin{equation}
\lvert\Psi_{\text{init}}\rangle:=\lvert\Gamma,Q,P\rangle.
\label{eq:init_psi_sampling}%
\end{equation}
%

\textit{Equidistant phase-space basis} functions are constructed as
\[
\phi_{\alpha}(q)=\langle{q}|\gamma,q_{i},p_{j}\rangle,\quad\alpha=iN_{p}+j,
\]
where
\begin{equation}%
\begin{split}
q_{i}  &  =Q+\frac{2i-(N_{q}-1)}{2}\Delta{}q,\quad0\leq{}i<N_{q},\\
p_{j}  &  =P+\frac{2j-(N_{p}-1)}{2}\Delta{}p,\quad0\leq{}j<N_{p}.
\end{split}
\label{eq:qipj}%
\end{equation}
Symbols $N_{q}$ and $N_{p}$ represent the numbers of points in the
corresponding phase-space coordinates. The size of basis is $N=N_{q}N_{p}$.

\begin{figure}
[htbp]\includegraphics[width=3.25in]{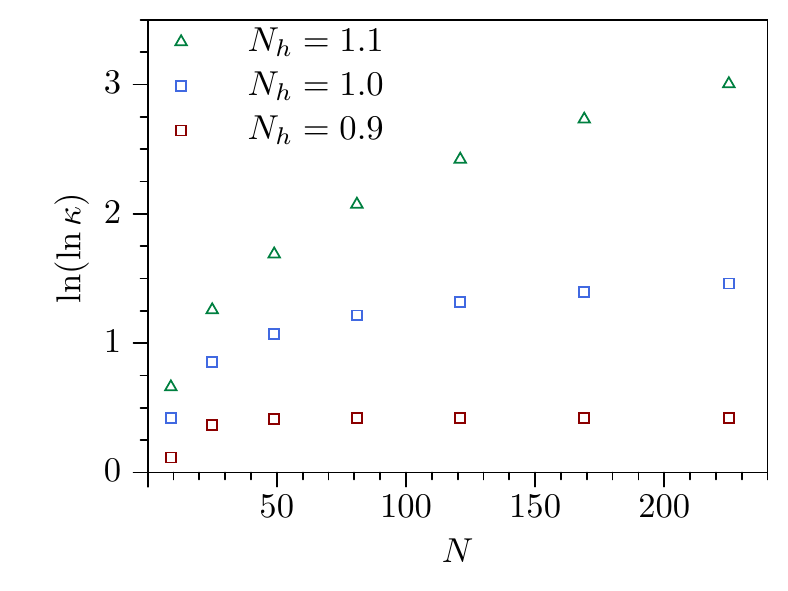}
\caption{Dependence of the condition number $\kappa$ of the
overlap matrix (\ref{eq:mat_S_el}) on the size $N$ and density of the basis constructed according
to Eq.~(\ref{eq:qipj}). $D=1$, $\gamma=1$, and $N_h$ denotes the number of basis functions
per phase-space area $h=2\pi \hbar$.} \label{fig:kappa}
\end{figure}

Grid spacings $\Delta{}q$ and $\Delta{}p$ are chosen to ensure a~given number
$N_{h}$ of basis functions per phase-space area $h=2\pi\hbar$ and a constant
absolute value of overlap between neighboring functions with the same index
$i$ or $j$. These requirements imply
\begin{equation}
\Delta{}q=\sqrt{\frac{2\pi}{N_{h}\gamma}}\,\text{ and }\,\Delta{}p=\hbar
\sqrt{\frac{2\pi\gamma}{N_{h}}}. \label{eq:dqdp}%
\end{equation}
Since the basis is nonorthogonal, ensuring fixed overlap between neighboring
functions does not guarantee constant linear independence of the basis with
increasing $N$, as illustrated in Fig.~\ref{fig:kappa}, which shows the
dependence of the condition number $\kappa$ of the overlap matrix $\mathbf{S}$
on $N$.

\textit{Monte Carlo\ basis }is generated with an
algorithm\cite{Kay_94c,Martinez_JPCA_00,DVS_pancakes} sampling the
coherent-state basis functions from the absolute value of the overlap
$\chi:=\langle\gamma,q,p|\Gamma,Q,P\rangle$ of the initial state
(\ref{eq:init_psi_sampling}) with a~basis state $\lvert\gamma,q,p\rangle$ of
Eq.~(\ref{eq:basis_fn}). The absolute value of this overlap, understood as
a~function of $q$ and $\,p$, is
\begin{equation}
\lvert\chi(q,\,p)\rvert^{2}\propto\exp\left[  -\frac{\gamma\Gamma}%
{\gamma+\Gamma}(q-Q)^{2}-\frac{1}{\hbar^{2}}\frac{(p-P)^{2}}{\gamma+\Gamma
}\right]  . \label{eq:chi_qp}%
\end{equation}
The overall procedure is as follows: \begin{algorithmic}[1]
\small
\Require{\Statex
\begin{itemize}[itemsep=0pt]
\item desired number of basis elements $N$
\item parameters $\xi>0$, $\gamma>0$, and $1>\epsilon>0$
\end{itemize}
}
\Ensure{Gaussian basis used in Eq.~(\ref{eq:psi})}
\State $\alpha:=1$
\While{$\alpha\leq{}N$}
\State sample $q,\,p$ from the distribution $\sim$ $\lvert\chi(q,\,p)\rvert^{2\xi}$
\State $\eta:=\text{sup}_{1\leq{}\beta<\alpha}\lvert\langle\phi_{\beta}|\gamma,q,p\rangle\rvert$
\If{$\eta<\epsilon$}
\State $\lvert\phi_{\alpha}\rangle:=\lvert\gamma,q,p\rangle$
\State $\alpha:=\alpha+1$
\EndIf
\EndWhile
\vspace*{0.5\baselineskip}
\end{algorithmic}The conditional statement in step~$5$ is added to improve the
condition number $\kappa$ of the resulting overlap matrix$~\mathbf{S}%
$.\cite{Martinez_ACP_02} A~modified approach based on orthogonal projections
was proposed\cite{Wu_03} by \citeauthor{Wu_03} in their Matching-Pursuit
Split-Operator Fourier-Transform technique.

%


\begin{thebibliography}{102}%
\makeatletter
\providecommand \@ifxundefined [1]{%
 \@ifx{#1\undefined}
}%
\providecommand \@ifnum [1]{%
 \ifnum #1\expandafter \@firstoftwo
 \else \expandafter \@secondoftwo
 \fi
}%
\providecommand \@ifx [1]{%
 \ifx #1\expandafter \@firstoftwo
 \else \expandafter \@secondoftwo
 \fi
}%
\providecommand \natexlab [1]{#1}%
\providecommand \enquote  [1]{``#1''}%
\providecommand \bibnamefont  [1]{#1}%
\providecommand \bibfnamefont [1]{#1}%
\providecommand \citenamefont [1]{#1}%
\providecommand \href@noop [0]{\@secondoftwo}%
\providecommand \href [0]{\begingroup \@sanitize@url \@href}%
\providecommand \@href[1]{\@@startlink{#1}\@@href}%
\providecommand \@@href[1]{\endgroup#1\@@endlink}%
\providecommand \@sanitize@url [0]{\catcode `\\12\catcode `\$12\catcode
  `\&12\catcode `\#12\catcode `\^12\catcode `\_12\catcode `\%12\relax}%
\providecommand \@@startlink[1]{}%
\providecommand \@@endlink[0]{}%
\providecommand \url  [0]{\begingroup\@sanitize@url \@url }%
\providecommand \@url [1]{\endgroup\@href {#1}{\urlprefix }}%
\providecommand \urlprefix  [0]{URL }%
\providecommand \Eprint [0]{\href }%
\providecommand \doibase [0]{http://dx.doi.org/}%
\providecommand \selectlanguage [0]{\@gobble}%
\providecommand \bibinfo  [0]{\@secondoftwo}%
\providecommand \bibfield  [0]{\@secondoftwo}%
\providecommand \translation [1]{[#1]}%
\providecommand \BibitemOpen [0]{}%
\providecommand \bibitemStop [0]{}%
\providecommand \bibitemNoStop [0]{.\EOS\space}%
\providecommand \EOS [0]{\spacefactor3000\relax}%
\providecommand \BibitemShut  [1]{\csname bibitem#1\endcsname}%
\let\auto@bib@innerbib\@empty
\bibitem [{\citenamefont {Bisgaard}\ \emph {et~al.}(2009)\citenamefont
  {Bisgaard}, \citenamefont {Clarkin}, \citenamefont {Wu}, \citenamefont {Lee},
  \citenamefont {Gessner}, \citenamefont {Hayden},\ and\ \citenamefont
  {Stolow}}]{Bisgaard2009}%
  \BibitemOpen
  \bibfield  {author} {\bibinfo {author} {\bibfnamefont {C.~Z.}\ \bibnamefont
  {Bisgaard}}, \bibinfo {author} {\bibfnamefont {O.~J.}\ \bibnamefont
  {Clarkin}}, \bibinfo {author} {\bibfnamefont {G.}~\bibnamefont {Wu}},
  \bibinfo {author} {\bibfnamefont {A.~M.~D.}\ \bibnamefont {Lee}}, \bibinfo
  {author} {\bibfnamefont {O.}~\bibnamefont {Gessner}}, \bibinfo {author}
  {\bibfnamefont {C.~C.}\ \bibnamefont {Hayden}}, \ and\ \bibinfo {author}
  {\bibfnamefont {A.}~\bibnamefont {Stolow}},\ }\href {\doibase
  10.1126/science.1169183} {\bibfield  {journal} {\bibinfo  {journal}
  {Science}\ }\textbf {\bibinfo {volume} {323}},\ \bibinfo {pages} {1464}
  (\bibinfo {year} {2009})}\BibitemShut {NoStop}%
\bibitem [{\citenamefont {Bressler}\ \emph {et~al.}(2009)\citenamefont
  {Bressler}, \citenamefont {Milne}, \citenamefont {Pham}, \citenamefont
  {ElNahhas}, \citenamefont {van~der Veen}, \citenamefont {Gawelda},
  \citenamefont {Johnson}, \citenamefont {Beaud}, \citenamefont {Grolimund},
  \citenamefont {Kaiser}, \citenamefont {Borca}, \citenamefont {Ingold},
  \citenamefont {Abela},\ and\ \citenamefont {Chergui}}]{Bressler2009}%
  \BibitemOpen
  \bibfield  {author} {\bibinfo {author} {\bibfnamefont {C.}~\bibnamefont
  {Bressler}}, \bibinfo {author} {\bibfnamefont {C.}~\bibnamefont {Milne}},
  \bibinfo {author} {\bibfnamefont {V.-T.}\ \bibnamefont {Pham}}, \bibinfo
  {author} {\bibfnamefont {A.}~\bibnamefont {ElNahhas}}, \bibinfo {author}
  {\bibfnamefont {R.~M.}\ \bibnamefont {van~der Veen}}, \bibinfo {author}
  {\bibfnamefont {W.}~\bibnamefont {Gawelda}}, \bibinfo {author} {\bibfnamefont
  {S.}~\bibnamefont {Johnson}}, \bibinfo {author} {\bibfnamefont
  {P.}~\bibnamefont {Beaud}}, \bibinfo {author} {\bibfnamefont
  {D.}~\bibnamefont {Grolimund}}, \bibinfo {author} {\bibfnamefont
  {M.}~\bibnamefont {Kaiser}}, \bibinfo {author} {\bibfnamefont {C.~N.}\
  \bibnamefont {Borca}}, \bibinfo {author} {\bibfnamefont {G.}~\bibnamefont
  {Ingold}}, \bibinfo {author} {\bibfnamefont {R.}~\bibnamefont {Abela}}, \
  and\ \bibinfo {author} {\bibfnamefont {M.}~\bibnamefont {Chergui}},\ }\href
  {\doibase 10.1126/science.1165733} {\bibfield  {journal} {\bibinfo  {journal}
  {Science}\ }\textbf {\bibinfo {volume} {323}},\ \bibinfo {pages} {489}
  (\bibinfo {year} {2009})}\BibitemShut {NoStop}%
\bibitem [{\citenamefont {Carbone}, \citenamefont {Kwon},\ and\ \citenamefont
  {Zewail}(2009)}]{Carbone2009}%
  \BibitemOpen
  \bibfield  {author} {\bibinfo {author} {\bibfnamefont {F.}~\bibnamefont
  {Carbone}}, \bibinfo {author} {\bibfnamefont {O.-H.}\ \bibnamefont {Kwon}}, \
  and\ \bibinfo {author} {\bibfnamefont {A.~H.}\ \bibnamefont {Zewail}},\
  }\href {\doibase 10.1126/science.1175005} {\bibfield  {journal} {\bibinfo
  {journal} {Science}\ }\textbf {\bibinfo {volume} {325}},\ \bibinfo {pages}
  {181} (\bibinfo {year} {2009})}\BibitemShut {NoStop}%
\bibitem [{\citenamefont {Miller}(2001)}]{miller:2001}%
  \BibitemOpen
  \bibfield  {author} {\bibinfo {author} {\bibfnamefont {W.~H.}\ \bibnamefont
  {Miller}},\ }\href {\doibase 10.1021/jp003712k} {\bibfield  {journal} {\bibinfo  {journal}
  {J.~Phys.\ Chem.~A}\ }\textbf {\bibinfo {volume} {105}},\ \bibinfo {pages}
  {2942} (\bibinfo {year} {2001})}\BibitemShut {NoStop}%
\bibitem [{\citenamefont {Herman}(1994)}]{Herman_94}%
  \BibitemOpen
  \bibfield  {author} {\bibinfo {author} {\bibfnamefont {M.~F.}\ \bibnamefont
  {Herman}},\ }\href {\doibase 10.1146/annurev.pc.45.100194.000503} {\bibfield
  {journal} {\bibinfo  {journal} {Annu.\ Rev.\ Phys.\ Chem.}\ }\textbf
  {\bibinfo {volume} {45}},\ \bibinfo {pages} {83} (\bibinfo {year}
  {1994})}\BibitemShut {NoStop}%
\bibitem [{\citenamefont {Thoss}\ and\ \citenamefont
  {Wang}(2004)}]{ThossWang_04}%
  \BibitemOpen
  \bibfield  {author} {\bibinfo {author} {\bibfnamefont {M.}~\bibnamefont
  {Thoss}}\ and\ \bibinfo {author} {\bibfnamefont {H.}~\bibnamefont {Wang}},\
  }\href {\doibase 10.1146/annurev.physchem.55.091602.094429} {\bibfield
  {journal} {\bibinfo  {journal} {Annu.\ Rev.\ Phys.\ Chem.}\ }\textbf
  {\bibinfo {volume} {55}},\ \bibinfo {pages} {299} (\bibinfo {year}
  {2004})}\BibitemShut {NoStop}%
\bibitem [{\citenamefont {Kay}(2005)}]{Kay_05}%
  \BibitemOpen
  \bibfield  {author} {\bibinfo {author} {\bibfnamefont {K.~G.}\ \bibnamefont
  {Kay}},\ }\href {\doibase 10.1146/annurev.physchem.56.092503.141257}
  {\bibfield  {journal} {\bibinfo  {journal} {Annu.\ Rev.\ Phys.\ Chem.}\
  }\textbf {\bibinfo {volume} {56}},\ \bibinfo {pages} {255} (\bibinfo {year}
  {2005})}\BibitemShut {NoStop}%
\bibitem [{\citenamefont {Meyer}, \citenamefont {Gatti},\ and\ \citenamefont
  {Worth}(2009)}]{book_MCTDH}%
  \BibitemOpen
  \bibinfo {editor} {\bibfnamefont {H.-D.}\ \bibnamefont {Meyer}}, \bibinfo
  {editor} {\bibfnamefont {F.}~\bibnamefont {Gatti}}, \ and\ \bibinfo {editor}
  {\bibfnamefont {G.~A.}\ \bibnamefont {Worth}},\ eds.,\ \href@noop {} {\emph
  {\bibinfo {title} {{Multidimensional Quantum Dynamics: MCTDH Theory and
  Applications}}}},\ \bibinfo {edition} {1st}\ ed.\ (\bibinfo  {publisher}
  {{Wiley-VCH}},\ \bibinfo {address} {Weinheim},\ \bibinfo {year}
  {2009})\BibitemShut {NoStop}%
\bibitem [{\citenamefont {Meyer}, \citenamefont {Manthe},\ and\ \citenamefont
  {Cederbaum}(1990)}]{Meyer:1990}%
  \BibitemOpen
  \bibfield  {author} {\bibinfo {author} {\bibfnamefont {H.-D.}\ \bibnamefont
  {Meyer}}, \bibinfo {author} {\bibfnamefont {U.}~\bibnamefont {Manthe}}, \
  and\ \bibinfo {author} {\bibfnamefont {L.}~\bibnamefont {Cederbaum}},\ }\href
  {\doibase 10.1016/0009-2614(90)87014-I} {\bibfield  {journal} {\bibinfo
  {journal} {Chem.\ Phys.\ Lett.}\ }\textbf {\bibinfo {volume} {165}},\
  \bibinfo {pages} {73} (\bibinfo {year} {1990})}\BibitemShut {NoStop}%
\bibitem [{\citenamefont {Burghardt}, \citenamefont {Meyer},\ and\
  \citenamefont {Cederbaum}(1999)}]{Burghardt_99}%
  \BibitemOpen
  \bibfield  {author} {\bibinfo {author} {\bibfnamefont {I.}~\bibnamefont
  {Burghardt}}, \bibinfo {author} {\bibfnamefont {H.-D.}\ \bibnamefont
  {Meyer}}, \ and\ \bibinfo {author} {\bibfnamefont {L.~S.}\ \bibnamefont
  {Cederbaum}},\ }\href {\doibase 10.1063/1.479574} {\bibfield  {journal}
  {\bibinfo  {journal} {J.~Chem.\ Phys.}\ }\textbf {\bibinfo {volume} {111}},\
  \bibinfo {pages} {2927} (\bibinfo {year} {1999})}\BibitemShut {NoStop}%
\bibitem [{\citenamefont {Ben-Nun}\ and\ \citenamefont
  {Mart\'{i}nez}(2002)}]{Martinez_ACP_02}%
  \BibitemOpen
  \bibfield  {author} {\bibinfo {author} {\bibfnamefont {M.}~\bibnamefont
  {Ben-Nun}}\ and\ \bibinfo {author} {\bibfnamefont {T.~J.}\ \bibnamefont
  {Mart\'{i}nez}},\ }\href {\doibase 10.1002/0471264318.ch7} {\bibfield
  {journal} {\bibinfo  {journal} {Adv.~Chem.~Phys.}\ }\textbf {\bibinfo
  {volume} {121}},\ \bibinfo {pages} {439} (\bibinfo {year}
  {2002})}\BibitemShut {NoStop}%
\bibitem [{\citenamefont {Tatchen}\ and\ \citenamefont
  {Pollak}(2009)}]{Tatchen_09}%
  \BibitemOpen
  \bibfield  {author} {\bibinfo {author} {\bibfnamefont {J.}~\bibnamefont
  {Tatchen}}\ and\ \bibinfo {author} {\bibfnamefont {E.}~\bibnamefont
  {Pollak}},\ }\href {\doibase 10.1063/1.3074100} {\bibfield  {journal}
  {\bibinfo  {journal} {J.~Chem.\ Phys.}\ }\textbf {\bibinfo {volume} {130}},\
  \bibinfo {eid} {041103} (\bibinfo {year} {2009})}\BibitemShut {NoStop}%
\bibitem [{\citenamefont {Ceotto}\ \emph {et~al.}(2009)\citenamefont {Ceotto},
  \citenamefont {Atahan}, \citenamefont {Tantardini},\ and\ \citenamefont
  {Aspuru-Guzik}}]{Ceotto_JCP_09}%
  \BibitemOpen
  \bibfield  {author} {\bibinfo {author} {\bibfnamefont {M.}~\bibnamefont
  {Ceotto}}, \bibinfo {author} {\bibfnamefont {S.}~\bibnamefont {Atahan}},
  \bibinfo {author} {\bibfnamefont {G.~F.}\ \bibnamefont {Tantardini}}, \ and\
  \bibinfo {author} {\bibfnamefont {A.}~\bibnamefont {Aspuru-Guzik}},\ }\href
  {\doibase 10.1063/1.3155062} {\bibfield  {journal} {\bibinfo  {journal}
  {J.~Chem.\ Phys.}\ }\textbf {\bibinfo {volume} {130}},\ \bibinfo {eid}
  {234113} (\bibinfo {year} {2009})}\BibitemShut {NoStop}%
\bibitem [{\citenamefont {Ceotto}, \citenamefont {Zhuang},\ and\ \citenamefont
  {Hase}(2013)}]{Ceotto_JCP_2013}%
  \BibitemOpen
  \bibfield  {author} {\bibinfo {author} {\bibfnamefont {M.}~\bibnamefont
  {Ceotto}}, \bibinfo {author} {\bibfnamefont {Y.}~\bibnamefont {Zhuang}}, \
  and\ \bibinfo {author} {\bibfnamefont {W.~L.}\ \bibnamefont {Hase}},\ }\href
  {\doibase 10.1063/1.4789759} {\bibfield  {journal} {\bibinfo  {journal}
  {J.~Chem.\ Phys.}\ }\textbf {\bibinfo {volume} {138}},\ \bibinfo {eid}
  {054116} (\bibinfo {year} {2013})}\BibitemShut {NoStop}%
\bibitem [{\citenamefont {Thompson}\ and\ \citenamefont
  {Mart\'{\i}nez}(2011)}]{Martinez_FD_11}%
  \BibitemOpen
  \bibfield  {author} {\bibinfo {author} {\bibfnamefont {A.~L.}\ \bibnamefont
  {Thompson}}\ and\ \bibinfo {author} {\bibfnamefont {T.~J.}\ \bibnamefont
  {Mart\'{\i}nez}},\ }\href {\doibase 10.1039/C1FD00003A} {\bibfield  {journal}
  {\bibinfo  {journal} {Faraday Discuss.}\ }\textbf {\bibinfo {volume} {150}},\
  \bibinfo {pages} {293} (\bibinfo {year} {2011})}\BibitemShut {NoStop}%
\bibitem [{\citenamefont {Worth}, \citenamefont {Robb},\ and\ \citenamefont
  {Burghardt}(2004)}]{Worth.Burghardt_FD_2004}%
  \BibitemOpen
  \bibfield  {author} {\bibinfo {author} {\bibfnamefont {G.~A.}\ \bibnamefont
  {Worth}}, \bibinfo {author} {\bibfnamefont {M.~A.}\ \bibnamefont {Robb}}, \
  and\ \bibinfo {author} {\bibfnamefont {I.}~\bibnamefont {Burghardt}},\ }\href
  {\doibase 10.1039/B314253A} {\bibfield  {journal} {\bibinfo  {journal}
  {Faraday Discuss.}\ }\textbf {\bibinfo {volume} {127}},\ \bibinfo {pages}
  {307} (\bibinfo {year} {2004})}\BibitemShut {NoStop}%
\bibitem [{\citenamefont {Worth}, \citenamefont {Robb},\ and\ \citenamefont
  {Lasorne}(2008)}]{Worth.Lasorne_MolPhys_2008}%
  \BibitemOpen
  \bibfield  {author} {\bibinfo {author} {\bibfnamefont {G.}~\bibnamefont
  {Worth}}, \bibinfo {author} {\bibfnamefont {M.}~\bibnamefont {Robb}}, \ and\
  \bibinfo {author} {\bibfnamefont {B.}~\bibnamefont {Lasorne}},\ }\href
  {\doibase 10.1080/00268970802172503} {\bibfield  {journal} {\bibinfo
  {journal} {Mol.\ Phys.}\ }\textbf {\bibinfo {volume} {106}},\ \bibinfo
  {pages} {2077} (\bibinfo {year} {2008})}\BibitemShut {NoStop}%
\bibitem [{\citenamefont {Saita}\ and\ \citenamefont
  {Shalashilin}(2012)}]{Saita.DVS_JCP_2012}%
  \BibitemOpen
  \bibfield  {author} {\bibinfo {author} {\bibfnamefont {K.}~\bibnamefont
  {Saita}}\ and\ \bibinfo {author} {\bibfnamefont {D.~V.}\ \bibnamefont
  {Shalashilin}},\ }\href {\doibase 10.1063/1.4734313} {\bibfield  {journal}
  {\bibinfo  {journal} {J.~Chem.\ Phys.}\ }\textbf {\bibinfo {volume} {137}},\
  \bibinfo {eid} {22A506} (\bibinfo {year} {2012})}\BibitemShut {NoStop}%
\bibitem [{\citenamefont {Van\'\i\v{c}ek}(2004)}]{Vanicek_04}%
  \BibitemOpen
  \bibfield  {author} {\bibinfo {author} {\bibfnamefont {J.}~\bibnamefont
  {Van\'\i\v{c}ek}},\ }\href {\doibase 10.1103/PhysRevE.70.055201} {\bibfield
  {journal} {\bibinfo  {journal} {Phys.\ Rev.~E}\ }\textbf {\bibinfo {volume}
  {70}},\ \bibinfo {pages} {055201} (\bibinfo {year} {2004})}\BibitemShut
  {NoStop}%
\bibitem [{\citenamefont {Van\'\i\v{c}ek}(2006)}]{Vanicek_06}%
  \BibitemOpen
  \bibfield  {author} {\bibinfo {author} {\bibfnamefont {J.}~\bibnamefont
  {Van\'\i\v{c}ek}},\ }\href {\doibase 10.1103/PhysRevE.73.046204} {\bibfield
  {journal} {\bibinfo  {journal} {Phys.\ Rev.~E}\ }\textbf {\bibinfo {volume}
  {73}},\ \bibinfo {pages} {046204} (\bibinfo {year} {2006})}\BibitemShut
  {NoStop}%
\bibitem [{\citenamefont {Wehrle}, \citenamefont {\v{S}ulc},\ and\
  \citenamefont {Van\'{\i}\v{c}ek}(2011)}]{Wehrle_11}%
  \BibitemOpen
  \bibfield  {author} {\bibinfo {author} {\bibfnamefont {M.}~\bibnamefont
  {Wehrle}}, \bibinfo {author} {\bibfnamefont {M.}~\bibnamefont {\v{S}ulc}}, \
  and\ \bibinfo {author} {\bibfnamefont {J.}~\bibnamefont {Van\'{\i}\v{c}ek}},\
  }\href {\doibase 10.2533/chimia.2011.334} {\bibfield  {journal} {\bibinfo
  {journal} {Chimia}\ }\textbf {\bibinfo {volume} {65}},\ \bibinfo {pages}
  {334} (\bibinfo {year} {2011})}\BibitemShut {NoStop}%
\bibitem [{\citenamefont {\v{S}ulc}\ and\ \citenamefont
  {Van\'\i\v{c}ek}(2012)}]{Sulc_MolPhys_12}%
  \BibitemOpen
  \bibfield  {author} {\bibinfo {author} {\bibfnamefont {M.}~\bibnamefont
  {\v{S}ulc}}\ and\ \bibinfo {author} {\bibfnamefont {J.}~\bibnamefont
  {Van\'\i\v{c}ek}},\ }\href {\doibase 10.1080/00268976.2012.668971} {\bibfield
   {journal} {\bibinfo  {journal} {Mol.\ Phys.}\ }\textbf {\bibinfo {volume}
  {110}},\ \bibinfo {pages} {945} (\bibinfo {year} {2012})}\BibitemShut
  {NoStop}%
\bibitem [{\citenamefont {Mukamel}(1982)}]{Mukamel_82}%
  \BibitemOpen
  \bibfield  {author} {\bibinfo {author} {\bibfnamefont {S.}~\bibnamefont
  {Mukamel}},\ }\href {\doibase 10.1063/1.443638} {\bibfield  {journal}
  {\bibinfo  {journal} {J.~Chem.\ Phys.}\ }\textbf {\bibinfo {volume} {77}},\
  \bibinfo {pages} {173} (\bibinfo {year} {1982})}\BibitemShut {NoStop}%
\bibitem [{\citenamefont {Mukamel}(1999)}]{book_Mukamel}%
  \BibitemOpen
  \bibfield  {author} {\bibinfo {author} {\bibfnamefont {S.}~\bibnamefont
  {Mukamel}},\ }\href@noop {} {\emph {\bibinfo {title} {{Principles of
  nonlinear optical spectroscopy}}}},\ \bibinfo {edition} {1st}\ ed.\ (\bibinfo
   {publisher} {Oxford University Press},\ \bibinfo {address} {New York},\
  \bibinfo {year} {1999})\BibitemShut {NoStop}%
\bibitem [{\citenamefont {Shi}\ and\ \citenamefont {Geva}(2005)}]{Shi_05}%
  \BibitemOpen
  \bibfield  {author} {\bibinfo {author} {\bibfnamefont {Q.}~\bibnamefont
  {Shi}}\ and\ \bibinfo {author} {\bibfnamefont {E.}~\bibnamefont {Geva}},\
  }\href {\doibase 10.1063/1.1843813} {\bibfield  {journal} {\bibinfo
  {journal} {J.~Chem.\ Phys.}\ }\textbf {\bibinfo {volume} {122}},\ \bibinfo
  {eid} {064506} (\bibinfo {year} {2005})}\BibitemShut {NoStop}%
\bibitem [{\citenamefont {McRobbie}\ \emph {et~al.}(2009)\citenamefont
  {McRobbie}, \citenamefont {Hanna}, \citenamefont {Shi},\ and\ \citenamefont
  {Geva}}]{Geva:2009}%
  \BibitemOpen
  \bibfield  {author} {\bibinfo {author} {\bibfnamefont {P.~L.}\ \bibnamefont
  {McRobbie}}, \bibinfo {author} {\bibfnamefont {G.}~\bibnamefont {Hanna}},
  \bibinfo {author} {\bibfnamefont {Q.}~\bibnamefont {Shi}}, \ and\ \bibinfo
  {author} {\bibfnamefont {E.}~\bibnamefont {Geva}},\ }\href {\doibase
  10.1021/ar800280s} {\bibfield  {journal} {\bibinfo  {journal}
  {Acc.~Chem.~Res.}\ }\textbf {\bibinfo {volume} {42}},\ \bibinfo {pages}
  {1299} (\bibinfo {year} {2009})}\BibitemShut {NoStop}%
\bibitem [{\citenamefont {Wang}, \citenamefont {Sun},\ and\ \citenamefont
  {Miller}(1998)}]{Miller_LSCIVR}%
  \BibitemOpen
  \bibfield  {author} {\bibinfo {author} {\bibfnamefont {H.}~\bibnamefont
  {Wang}}, \bibinfo {author} {\bibfnamefont {X.}~\bibnamefont {Sun}}, \ and\
  \bibinfo {author} {\bibfnamefont {W.~H.}\ \bibnamefont {Miller}},\ }\href
  {\doibase 10.1063/1.476447} {\bibfield  {journal} {\bibinfo  {journal}
  {J.~Chem.\ Phys.}\ }\textbf {\bibinfo {volume} {108}},\ \bibinfo {pages}
  {9726} (\bibinfo {year} {1998})}\BibitemShut {NoStop}%
\bibitem [{note2()}]{note2}%
  \BibitemOpen
  \bibinfo {note} {Some authors\protect\cite{Shi_05,Geva:2009} refer to the DR
  as the LSC-IVR since the DR can be understood as a linearized semiclassical
  approximation for fidelity amplitude.\protect\cite{Shi_05,Vanicek_06}
  However, in the notation of Sec.~\protect\ref{sec:theory}, LSC-IVR usually
  refers to the approximation\protect\cite{Miller_LSCIVR}
  $\operatorname{Tr}(\hat{A}(0)\hat{B}(t))\approx h^{-D}\int
  dxA_{W}(x^{0})B_{W}(x^{t})$, which can be also used to approximate fidelity
  (i.e., fidelity amplitude squared),\protect\cite{Vanicek_06} but does not
  include any quantum dynamical effects; the only quantum effects are static,
  arising from replacing classical observables by their Wigner transforms.
  Therefore we prefer avoiding the name LSC-IVR when referring to the DR of
  fidelity amplitude, since the DR does incorporate dynamical interference
  effects even when the description based on the LSC-IVR of fidelity is purely
  classical.}\BibitemShut {Stop}%
\bibitem [{\citenamefont {Li}, \citenamefont {Fang},\ and\ \citenamefont
  {Martens}(1996)}]{Li_96}%
  \BibitemOpen
  \bibfield  {author} {\bibinfo {author} {\bibfnamefont {Z.}~\bibnamefont
  {Li}}, \bibinfo {author} {\bibfnamefont {J.-Y.}\ \bibnamefont {Fang}}, \ and\
  \bibinfo {author} {\bibfnamefont {C.~C.}\ \bibnamefont {Martens}},\ }\href
  {\doibase 10.1063/1.471407} {\bibfield  {journal} {\bibinfo  {journal}
  {J.~Chem.\ Phys.}\ }\textbf {\bibinfo {volume} {104}},\ \bibinfo {pages}
  {6919} (\bibinfo {year} {1996})}\BibitemShut {NoStop}%
\bibitem [{\citenamefont {Egorov}, \citenamefont {Rabani},\ and\ \citenamefont
  {Berne}(1998)}]{Egorov_98}%
  \BibitemOpen
  \bibfield  {author} {\bibinfo {author} {\bibfnamefont {S.~A.}\ \bibnamefont
  {Egorov}}, \bibinfo {author} {\bibfnamefont {E.}~\bibnamefont {Rabani}}, \
  and\ \bibinfo {author} {\bibfnamefont {B.~J.}\ \bibnamefont {Berne}},\ }\href
  {\doibase 10.1063/1.475512} {\bibfield  {journal} {\bibinfo  {journal}
  {J.~Chem.\ Phys.}\ }\textbf {\bibinfo {volume} {108}},\ \bibinfo {pages}
  {1407} (\bibinfo {year} {1998})}\BibitemShut {NoStop}%
\bibitem [{\citenamefont {Egorov}, \citenamefont {Rabani},\ and\ \citenamefont
  {Berne}(1999)}]{Egorov_99}%
  \BibitemOpen
  \bibfield  {author} {\bibinfo {author} {\bibfnamefont {S.~A.}\ \bibnamefont
  {Egorov}}, \bibinfo {author} {\bibfnamefont {E.}~\bibnamefont {Rabani}}, \
  and\ \bibinfo {author} {\bibfnamefont {B.~J.}\ \bibnamefont {Berne}},\ }\href
  {\doibase 10.1063/1.478420} {\bibfield  {journal} {\bibinfo  {journal}
  {J.~Chem.\ Phys.}\ }\textbf {\bibinfo {volume} {110}},\ \bibinfo {pages}
  {5238} (\bibinfo {year} {1999})}\BibitemShut {NoStop}%
\bibitem [{\citenamefont {Shemetulskis}\ and\ \citenamefont
  {Loring}(1992)}]{Shemetulskis_92}%
  \BibitemOpen
  \bibfield  {author} {\bibinfo {author} {\bibfnamefont {N.~E.}\ \bibnamefont
  {Shemetulskis}}\ and\ \bibinfo {author} {\bibfnamefont {R.~F.}\ \bibnamefont
  {Loring}},\ }\href {\doibase 10.1063/1.463248} {\bibfield  {journal}
  {\bibinfo  {journal} {J.~Chem.\ Phys.}\ }\textbf {\bibinfo {volume} {97}},\
  \bibinfo {pages} {1217} (\bibinfo {year} {1992})}\BibitemShut {NoStop}%
\bibitem [{\citenamefont {Rost}(1995)}]{Rost_95}%
  \BibitemOpen
  \bibfield  {author} {\bibinfo {author} {\bibfnamefont {J.~M.}\ \bibnamefont
  {Rost}},\ }\href {http://stacks.iop.org/0953-4075/28/i=19/a=002} {\bibfield
  {journal} {\bibinfo  {journal} {J.~Phys.~B}\ }\textbf {\bibinfo {volume}
  {28}},\ \bibinfo {pages} {L601} (\bibinfo {year} {1995})}\BibitemShut
  {NoStop}%
\bibitem [{\citenamefont {Zimmermann}\ and\ \citenamefont
  {Van\'{\i}\v{c}ek}(2012{\natexlab{a}})}]{Zimmermann_12}%
  \BibitemOpen
  \bibfield  {author} {\bibinfo {author} {\bibfnamefont {T.}~\bibnamefont
  {Zimmermann}}\ and\ \bibinfo {author} {\bibfnamefont {J.}~\bibnamefont
  {Van\'{\i}\v{c}ek}},\ }\href {\doibase 10.1063/1.3690458} {\bibfield
  {journal} {\bibinfo  {journal} {J.~Chem.\ Phys.}\ }\textbf {\bibinfo {volume}
  {136}},\ \bibinfo {eid} {094106} (\bibinfo {year}
  {2012}{\natexlab{a}})}\BibitemShut {NoStop}%
\bibitem [{\citenamefont {Zimmermann}\ and\ \citenamefont
  {Van\'{\i}\v{c}ek}(2012{\natexlab{b}})}]{Zimmermann_12b}%
  \BibitemOpen
  \bibfield  {author} {\bibinfo {author} {\bibfnamefont {T.}~\bibnamefont
  {Zimmermann}}\ and\ \bibinfo {author} {\bibfnamefont {J.}~\bibnamefont
  {Van\'{\i}\v{c}ek}},\ }\href {\doibase 10.1063/1.4738878} {\bibfield
  {journal} {\bibinfo  {journal} {J.~Chem.\ Phys.}\ }\textbf {\bibinfo {volume}
  {137}},\ \bibinfo {eid} {22A516} (\bibinfo {year}
  {2012}{\natexlab{b}})}\BibitemShut {NoStop}%
\bibitem [{\citenamefont {Petitjean}\ \emph {et~al.}(2007)\citenamefont
  {Petitjean}, \citenamefont {Bevilaqua}, \citenamefont {Heller},\ and\
  \citenamefont {Jacquod}}]{Petitjean_07}%
  \BibitemOpen
  \bibfield  {author} {\bibinfo {author} {\bibfnamefont {C.}~\bibnamefont
  {Petitjean}}, \bibinfo {author} {\bibfnamefont {D.~V.}\ \bibnamefont
  {Bevilaqua}}, \bibinfo {author} {\bibfnamefont {E.~J.}\ \bibnamefont
  {Heller}}, \ and\ \bibinfo {author} {\bibfnamefont {P.}~\bibnamefont
  {Jacquod}},\ }\href {\doibase 10.1103/PhysRevLett.98.164101} {\bibfield
  {journal} {\bibinfo  {journal} {Phys.\ Rev.\ Lett.}\ }\textbf {\bibinfo
  {volume} {98}},\ \bibinfo {pages} {164101} (\bibinfo {year}
  {2007})}\BibitemShut {NoStop}%
\bibitem [{\citenamefont {Zimmermann}\ and\ \citenamefont
  {Van\'{\i}\v{c}ek}(2010)}]{Zimmermann_10}%
  \BibitemOpen
  \bibfield  {author} {\bibinfo {author} {\bibfnamefont {T.}~\bibnamefont
  {Zimmermann}}\ and\ \bibinfo {author} {\bibfnamefont {J.}~\bibnamefont
  {Van\'{\i}\v{c}ek}},\ }\href {\doibase 10.1063/1.3451266} {\bibfield
  {journal} {\bibinfo  {journal} {J.~Chem.\ Phys.}\ }\textbf {\bibinfo {volume}
  {132}},\ \bibinfo {eid} {241101} (\bibinfo {year} {2010})}\BibitemShut
  {NoStop}%
\bibitem [{\citenamefont {Li}, \citenamefont {Mollica},\ and\ \citenamefont
  {Van\'{\i}\v{c}ek}(2009)}]{Li_09}%
  \BibitemOpen
  \bibfield  {author} {\bibinfo {author} {\bibfnamefont {B.}~\bibnamefont
  {Li}}, \bibinfo {author} {\bibfnamefont {C.}~\bibnamefont {Mollica}}, \ and\
  \bibinfo {author} {\bibfnamefont {J.}~\bibnamefont {Van\'{\i}\v{c}ek}},\
  }\href {\doibase 10.1063/1.3187240} {\bibfield  {journal} {\bibinfo
  {journal} {J.~Chem.\ Phys.}\ }\textbf {\bibinfo {volume} {131}},\ \bibinfo
  {eid} {041101} (\bibinfo {year} {2009})}\BibitemShut {NoStop}%
\bibitem [{\citenamefont {Zimmermann}\ \emph {et~al.}(2010)\citenamefont
  {Zimmermann}, \citenamefont {Ruppen}, \citenamefont {Li},\ and\ \citenamefont
  {Van\'{\i}\v{c}ek}}]{Zimmermann_IJQC_10}%
  \BibitemOpen
  \bibfield  {author} {\bibinfo {author} {\bibfnamefont {T.}~\bibnamefont
  {Zimmermann}}, \bibinfo {author} {\bibfnamefont {J.}~\bibnamefont {Ruppen}},
  \bibinfo {author} {\bibfnamefont {B.}~\bibnamefont {Li}}, \ and\ \bibinfo
  {author} {\bibfnamefont {J.}~\bibnamefont {Van\'{\i}\v{c}ek}},\ }\href
  {\doibase 10.1002/qua.22730} {\bibfield  {journal} {\bibinfo  {journal}
  {Int.\ J.~Quant.\ Chem.}\ }\textbf {\bibinfo {volume} {110}},\ \bibinfo
  {pages} {2426} (\bibinfo {year} {2010})}\BibitemShut {NoStop}%
\bibitem [{\citenamefont {Wang}\ \emph {et~al.}(2005)\citenamefont {Wang},
  \citenamefont {Casati}, \citenamefont {Li},\ and\ \citenamefont
  {Prosen}}]{Wang_05}%
  \BibitemOpen
  \bibfield  {author} {\bibinfo {author} {\bibfnamefont {W.}~\bibnamefont
  {Wang}}, \bibinfo {author} {\bibfnamefont {G.}~\bibnamefont {Casati}},
  \bibinfo {author} {\bibfnamefont {B.}~\bibnamefont {Li}}, \ and\ \bibinfo
  {author} {\bibfnamefont {T.}~\bibnamefont {Prosen}},\ }\href {\doibase
  10.1103/PhysRevE.71.037202} {\bibfield  {journal} {\bibinfo  {journal}
  {Phys.\ Rev.~E}\ }\textbf {\bibinfo {volume} {71}},\ \bibinfo {pages}
  {037202} (\bibinfo {year} {2005})}\BibitemShut {NoStop}%
\bibitem [{\citenamefont {Ares}\ and\ \citenamefont
  {Wisniacki}(2009)}]{Ares_09}%
  \BibitemOpen
  \bibfield  {author} {\bibinfo {author} {\bibfnamefont {N.}~\bibnamefont
  {Ares}}\ and\ \bibinfo {author} {\bibfnamefont {D.~A.}\ \bibnamefont
  {Wisniacki}},\ }\href {\doibase 10.1103/PhysRevE.80.046216} {\bibfield
  {journal} {\bibinfo  {journal} {Phys.\ Rev.~E}\ }\textbf {\bibinfo {volume}
  {80}},\ \bibinfo {pages} {046216} (\bibinfo {year} {2009})}\BibitemShut
  {NoStop}%
\bibitem [{\citenamefont {Wisniacki}, \citenamefont {Ares},\ and\ \citenamefont
  {Vergini}(2010)}]{Wisniacki_10}%
  \BibitemOpen
  \bibfield  {author} {\bibinfo {author} {\bibfnamefont {D.~A.}\ \bibnamefont
  {Wisniacki}}, \bibinfo {author} {\bibfnamefont {N.}~\bibnamefont {Ares}}, \
  and\ \bibinfo {author} {\bibfnamefont {E.~G.}\ \bibnamefont {Vergini}},\
  }\href {\doibase 10.1103/PhysRevLett.104.254101} {\bibfield  {journal}
  {\bibinfo  {journal} {Phys.\ Rev.\ Lett.}\ }\textbf {\bibinfo {volume}
  {104}},\ \bibinfo {pages} {254101} (\bibinfo {year} {2010})}\BibitemShut
  {NoStop}%
\bibitem [{\citenamefont {Garc\'{i}a-Mata}\ and\ \citenamefont
  {Wisniacki}(2011)}]{Wisniacki_11}%
  \BibitemOpen
  \bibfield  {author} {\bibinfo {author} {\bibfnamefont {I.}~\bibnamefont
  {Garc\'{i}a-Mata}}\ and\ \bibinfo {author} {\bibfnamefont {D.~A.}\
  \bibnamefont {Wisniacki}},\ }\href
  {http://stacks.iop.org/1751-8121/44/i=31/a=315101} {\bibfield  {journal}
  {\bibinfo  {journal} {J.~Phys.~A}\ }\textbf {\bibinfo {volume} {44}},\
  \bibinfo {pages} {315101} (\bibinfo {year} {2011})}\BibitemShut {NoStop}%
\bibitem [{\citenamefont {Thompson}\ and\ \citenamefont
  {Makri}(1999)}]{Thompson_Makri:1999}%
  \BibitemOpen
  \bibfield  {author} {\bibinfo {author} {\bibfnamefont {K.}~\bibnamefont
  {Thompson}}\ and\ \bibinfo {author} {\bibfnamefont {N.}~\bibnamefont
  {Makri}},\ }\href {\doibase 10.1103/PhysRevE.59.R4729} {\bibfield  {journal}
  {\bibinfo  {journal} {Phys.\ Rev.~E}\ }\textbf {\bibinfo {volume} {59}},\
  \bibinfo {pages} {R4729} (\bibinfo {year} {1999})}\BibitemShut {NoStop}%
\bibitem [{\citenamefont {K\"{u}hn}\ and\ \citenamefont
  {Makri}(1999)}]{Kuhn_Makri:1999}%
  \BibitemOpen
  \bibfield  {author} {\bibinfo {author} {\bibfnamefont {O.}~\bibnamefont
  {K\"{u}hn}}\ and\ \bibinfo {author} {\bibfnamefont {N.}~\bibnamefont
  {Makri}},\ }\href {\doibase 10.1021/jp991836v} {\bibfield  {journal}
  {\bibinfo  {journal} {J.~Phys.\ Chem.~A}\ }\textbf {\bibinfo {volume}
  {103}},\ \bibinfo {pages} {9487} (\bibinfo {year} {1999})}\BibitemShut
  {NoStop}%
\bibitem [{\citenamefont {Mollica}\ and\ \citenamefont
  {Van\'{\i}\v{c}ek}(2011)}]{Mollica_PRL_11}%
  \BibitemOpen
  \bibfield  {author} {\bibinfo {author} {\bibfnamefont {C.}~\bibnamefont
  {Mollica}}\ and\ \bibinfo {author} {\bibfnamefont {J.}~\bibnamefont
  {Van\'{\i}\v{c}ek}},\ }\href {\doibase 10.1103/PhysRevLett.107.214101}
  {\bibfield  {journal} {\bibinfo  {journal} {Phys.\ Rev.\ Lett.}\ }\textbf
  {\bibinfo {volume} {107}},\ \bibinfo {pages} {214101} (\bibinfo {year}
  {2011})}\BibitemShut {NoStop}%
\bibitem [{\citenamefont {Heller}(1991)}]{Heller_91}%
  \BibitemOpen
  \bibfield  {author} {\bibinfo {author} {\bibfnamefont {E.~J.}\ \bibnamefont
  {Heller}},\ }\href {\doibase 10.1063/1.459848} {\bibfield  {journal}
  {\bibinfo  {journal} {J.~Chem.\ Phys.}\ }\textbf {\bibinfo {volume} {94}},\
  \bibinfo {pages} {2723} (\bibinfo {year} {1991})}\BibitemShut {NoStop}%
\bibitem [{\citenamefont {Zambrano}\ and\ \citenamefont {Ozorio~de
  Almeida}(2011)}]{Zambrano_PRE_11}%
  \BibitemOpen
  \bibfield  {author} {\bibinfo {author} {\bibfnamefont {E.}~\bibnamefont
  {Zambrano}}\ and\ \bibinfo {author} {\bibfnamefont {A.~M.}\ \bibnamefont
  {Ozorio~de Almeida}},\ }\href {\doibase 10.1103/PhysRevE.84.045201}
  {\bibfield  {journal} {\bibinfo  {journal} {Phys.\ Rev.~E}\ }\textbf
  {\bibinfo {volume} {84}},\ \bibinfo {pages} {045201(R)} (\bibinfo {year}
  {2011})}\BibitemShut {NoStop}%
\bibitem [{\citenamefont {Burghardt}, \citenamefont {Giri},\ and\ \citenamefont
  {Worth}(2008)}]{Burghardt:2008}%
  \BibitemOpen
  \bibfield  {author} {\bibinfo {author} {\bibfnamefont {I.}~\bibnamefont
  {Burghardt}}, \bibinfo {author} {\bibfnamefont {K.}~\bibnamefont {Giri}}, \
  and\ \bibinfo {author} {\bibfnamefont {G.~A.}\ \bibnamefont {Worth}},\ }\href
  {\doibase 10.1063/1.2996349} {\bibfield  {journal} {\bibinfo  {journal}
  {J.~Chem.\ Phys.}\ }\textbf {\bibinfo {volume} {129}},\ \bibinfo {eid}
  {174104} (\bibinfo {year} {2008})}\BibitemShut {NoStop}%
\bibitem [{\citenamefont {Mart\'{\i}nez}, \citenamefont {Ben-Nun},\ and\
  \citenamefont {Ashkenazi}(1996)}]{Martinez_JCP_96}%
  \BibitemOpen
  \bibfield  {author} {\bibinfo {author} {\bibfnamefont {T.~J.}\ \bibnamefont
  {Mart\'{\i}nez}}, \bibinfo {author} {\bibfnamefont {M.}~\bibnamefont
  {Ben-Nun}}, \ and\ \bibinfo {author} {\bibfnamefont {G.}~\bibnamefont
  {Ashkenazi}},\ }\href {\doibase 10.1063/1.471108} {\bibfield  {journal}
  {\bibinfo  {journal} {J.~Chem.\ Phys.}\ }\textbf {\bibinfo {volume} {104}},\
  \bibinfo {pages} {2847} (\bibinfo {year} {1996})}\BibitemShut {NoStop}%
\bibitem [{\citenamefont {Mart\'{\i}nez}, \citenamefont {Ben-Nun},\ and\
  \citenamefont {Levine}(1996)}]{Martinez_JPC_96}%
  \BibitemOpen
  \bibfield  {author} {\bibinfo {author} {\bibfnamefont {T.~J.}\ \bibnamefont
  {Mart\'{\i}nez}}, \bibinfo {author} {\bibfnamefont {M.}~\bibnamefont
  {Ben-Nun}}, \ and\ \bibinfo {author} {\bibfnamefont {R.~D.}\ \bibnamefont
  {Levine}},\ }\href {\doibase 10.1021/jp953105a} {\bibfield  {journal}
  {\bibinfo  {journal} {J.~Phys.\ Chem.}\ }\textbf {\bibinfo {volume} {100}},\
  \bibinfo {pages} {7884} (\bibinfo {year} {1996})}\BibitemShut {NoStop}%
\bibitem [{\citenamefont {Sawada}\ \emph {et~al.}(1985)\citenamefont {Sawada},
  \citenamefont {Heather}, \citenamefont {Jackson},\ and\ \citenamefont
  {Metiu}}]{Sawada_85}%
  \BibitemOpen
  \bibfield  {author} {\bibinfo {author} {\bibfnamefont {S.-I.}\ \bibnamefont
  {Sawada}}, \bibinfo {author} {\bibfnamefont {R.}~\bibnamefont {Heather}},
  \bibinfo {author} {\bibfnamefont {B.}~\bibnamefont {Jackson}}, \ and\
  \bibinfo {author} {\bibfnamefont {H.}~\bibnamefont {Metiu}},\ }\href
  {\doibase 10.1063/1.449204} {\bibfield  {journal} {\bibinfo  {journal}
  {J.~Chem.\ Phys.}\ }\textbf {\bibinfo {volume} {83}},\ \bibinfo {pages}
  {3009} (\bibinfo {year} {1985})}\BibitemShut {NoStop}%
\bibitem [{\citenamefont {Heather}\ and\ \citenamefont
  {Metiu}(1985)}]{Metiu_CPL_85}%
  \BibitemOpen
  \bibfield  {author} {\bibinfo {author} {\bibfnamefont {R.}~\bibnamefont
  {Heather}}\ and\ \bibinfo {author} {\bibfnamefont {H.}~\bibnamefont
  {Metiu}},\ }\href {\doibase 10.1016/0009-2614(85)85353-7} {\bibfield
  {journal} {\bibinfo  {journal} {Chem.\ Phys.\ Lett.}\ }\textbf {\bibinfo
  {volume} {118}},\ \bibinfo {pages} {558} (\bibinfo {year}
  {1985})}\BibitemShut {NoStop}%
\bibitem [{\citenamefont {Heather}\ and\ \citenamefont
  {Metiu}(1986)}]{Metiu_JCP_86}%
  \BibitemOpen
  \bibfield  {author} {\bibinfo {author} {\bibfnamefont {R.}~\bibnamefont
  {Heather}}\ and\ \bibinfo {author} {\bibfnamefont {H.}~\bibnamefont
  {Metiu}},\ }\href {\doibase 10.1063/1.450255} {\bibfield  {journal} {\bibinfo
   {journal} {J.~Chem.\ Phys.}\ }\textbf {\bibinfo {volume} {84}},\ \bibinfo
  {pages} {3250} (\bibinfo {year} {1986})}\BibitemShut {NoStop}%
\bibitem [{\citenamefont {Shalashilin}\ and\ \citenamefont
  {Child}(2000)}]{Shalashilin_JCP_00}%
  \BibitemOpen
  \bibfield  {author} {\bibinfo {author} {\bibfnamefont {D.~V.}\ \bibnamefont
  {Shalashilin}}\ and\ \bibinfo {author} {\bibfnamefont {M.~S.}\ \bibnamefont
  {Child}},\ }\href {\doibase 10.1063/1.1322075} {\bibfield  {journal}
  {\bibinfo  {journal} {J.~Chem.\ Phys.}\ }\textbf {\bibinfo {volume} {113}},\
  \bibinfo {pages} {10028} (\bibinfo {year} {2000})}\BibitemShut {NoStop}%
\bibitem [{\citenamefont {Shalashilin}\ and\ \citenamefont
  {Child}(2001{\natexlab{a}})}]{Shalashilin_JCP_01a}%
  \BibitemOpen
  \bibfield  {author} {\bibinfo {author} {\bibfnamefont {D.~V.}\ \bibnamefont
  {Shalashilin}}\ and\ \bibinfo {author} {\bibfnamefont {M.~S.}\ \bibnamefont
  {Child}},\ }\href {\doibase 10.1063/1.1367392} {\bibfield  {journal}
  {\bibinfo  {journal} {J.~Chem.\ Phys.}\ }\textbf {\bibinfo {volume} {114}},\
  \bibinfo {pages} {9296} (\bibinfo {year} {2001}{\natexlab{a}})}\BibitemShut
  {NoStop}%
\bibitem [{\citenamefont {Shalashilin}\ and\ \citenamefont
  {Child}(2001{\natexlab{b}})}]{Shalashilin_JCP_01b}%
  \BibitemOpen
  \bibfield  {author} {\bibinfo {author} {\bibfnamefont {D.~V.}\ \bibnamefont
  {Shalashilin}}\ and\ \bibinfo {author} {\bibfnamefont {M.~S.}\ \bibnamefont
  {Child}},\ }\href {\doibase 10.1063/1.1394939} {\bibfield  {journal}
  {\bibinfo  {journal} {J.~Chem.\ Phys.}\ }\textbf {\bibinfo {volume} {115}},\
  \bibinfo {pages} {5367} (\bibinfo {year} {2001}{\natexlab{b}})}\BibitemShut
  {NoStop}%
\bibitem [{\citenamefont {Shalashilin}\ and\ \citenamefont
  {Child}(2004)}]{Shalashilin_ChemPhys_04}%
  \BibitemOpen
  \bibfield  {author} {\bibinfo {author} {\bibfnamefont {D.~V.}\ \bibnamefont
  {Shalashilin}}\ and\ \bibinfo {author} {\bibfnamefont {M.~S.}\ \bibnamefont
  {Child}},\ }\href {\doibase 10.1016/j.chemphys.2004.06.013} {\bibfield
  {journal} {\bibinfo  {journal} {Chem.\ Phys.}\ }\textbf {\bibinfo {volume}
  {304}},\ \bibinfo {pages} {103} (\bibinfo {year} {2004})}\BibitemShut
  {NoStop}%
\bibitem [{\citenamefont {Pollard}, \citenamefont {Lee},\ and\ \citenamefont
  {Mathies}(1990)}]{Pollard_Mathies:1990}%
  \BibitemOpen
  \bibfield  {author} {\bibinfo {author} {\bibfnamefont {W.~T.}\ \bibnamefont
  {Pollard}}, \bibinfo {author} {\bibfnamefont {S.-Y.}\ \bibnamefont {Lee}}, \
  and\ \bibinfo {author} {\bibfnamefont {R.~A.}\ \bibnamefont {Mathies}},\
  }\href {\doibase 10.1063/1.457815} {\bibfield  {journal} {\bibinfo  {journal}
  {J.~Chem.\ Phys.}\ }\textbf {\bibinfo {volume} {92}},\ \bibinfo {pages}
  {4012} (\bibinfo {year} {1990})}\BibitemShut {NoStop}%
\bibitem [{\citenamefont {Tannor}(2004)}]{book_Tannor}%
  \BibitemOpen
  \bibfield  {author} {\bibinfo {author} {\bibfnamefont {D.~J.}\ \bibnamefont
  {Tannor}},\ }\href@noop {} {\emph {\bibinfo {title} {Introduction to Quantum
  Mechanics: A Time-Dependent Perspective}}}\ (\bibinfo  {publisher}
  {University Science Books},\ \bibinfo {address} {Sausalito, California},\
  \bibinfo {year} {2004})\BibitemShut {NoStop}%
\bibitem [{\citenamefont {Gorin}\ \emph {et~al.}(2006)\citenamefont {Gorin},
  \citenamefont {Prosen}, \citenamefont {Seligman},\ and\ \citenamefont
  {\v{Z}nidari\v{c}}}]{Gorin_06}%
  \BibitemOpen
  \bibfield  {author} {\bibinfo {author} {\bibfnamefont {T.}~\bibnamefont
  {Gorin}}, \bibinfo {author} {\bibfnamefont {T.}~\bibnamefont {Prosen}},
  \bibinfo {author} {\bibfnamefont {T.~H.}\ \bibnamefont {Seligman}}, \ and\
  \bibinfo {author} {\bibfnamefont {M.}~\bibnamefont {\v{Z}nidari\v{c}}},\
  }\href {\doibase 10.1016/j.physrep.2006.09.003} {\bibfield  {journal}
  {\bibinfo  {journal} {Phys.~Rep.}\ }\textbf {\bibinfo {volume} {435}},\
  \bibinfo {pages} {33} (\bibinfo {year} {2006})}\BibitemShut {NoStop}%
\bibitem [{\citenamefont {Jacquod}\ and\ \citenamefont
  {Petitjean}(2009)}]{Jacquod_09}%
  \BibitemOpen
  \bibfield  {author} {\bibinfo {author} {\bibfnamefont {P.}~\bibnamefont
  {Jacquod}}\ and\ \bibinfo {author} {\bibfnamefont {C.}~\bibnamefont
  {Petitjean}},\ }\href {\doibase 10.1080/00018730902831009} {\bibfield
  {journal} {\bibinfo  {journal} {Adv.~Phys.}\ }\textbf {\bibinfo {volume}
  {58}},\ \bibinfo {pages} {67} (\bibinfo {year} {2009})}\BibitemShut {NoStop}%
\bibitem [{\citenamefont {Pastawski}\ \emph {et~al.}(2000)\citenamefont
  {Pastawski}, \citenamefont {Levstein}, \citenamefont {G.}, \citenamefont
  {J.},\ and\ \citenamefont {J.}}]{Pastawski_00}%
  \BibitemOpen
  \bibfield  {author} {\bibinfo {author} {\bibfnamefont {H.~M.}\ \bibnamefont
  {Pastawski}}, \bibinfo {author} {\bibfnamefont {P.~R.}\ \bibnamefont
  {Levstein}}, \bibinfo {author} {\bibfnamefont {U.}~\bibnamefont {G.}},
  \bibinfo {author} {\bibfnamefont {R.}~\bibnamefont {J.}}, \ and\ \bibinfo
  {author} {\bibfnamefont {H.}~\bibnamefont {J.}},\ }\href {\doibase
  10.1016/S0378-4371(00)00146-1} {\bibfield  {journal} {\bibinfo  {journal}
  {Physica~A}\ }\textbf {\bibinfo {volume} {283}},\ \bibinfo {pages} {166}
  (\bibinfo {year} {2000})}\BibitemShut {NoStop}%
\bibitem [{\citenamefont {Cucchietti}\ \emph {et~al.}(2003)\citenamefont
  {Cucchietti}, \citenamefont {Dalvit}, \citenamefont {Paz},\ and\
  \citenamefont {Zurek}}]{Cucchietti_03}%
  \BibitemOpen
  \bibfield  {author} {\bibinfo {author} {\bibfnamefont {F.~M.}\ \bibnamefont
  {Cucchietti}}, \bibinfo {author} {\bibfnamefont {D.~A.~R.}\ \bibnamefont
  {Dalvit}}, \bibinfo {author} {\bibfnamefont {J.~P.}\ \bibnamefont {Paz}}, \
  and\ \bibinfo {author} {\bibfnamefont {W.~H.}\ \bibnamefont {Zurek}},\ }\href
  {\doibase 10.1103/PhysRevLett.91.210403} {\bibfield  {journal} {\bibinfo
  {journal} {Phys.\ Rev.\ Lett.}\ }\textbf {\bibinfo {volume} {91}},\ \bibinfo
  {pages} {210403} (\bibinfo {year} {2003})}\BibitemShut {NoStop}%
\bibitem [{\citenamefont {Van\'{\i}\v{c}ek}\ and\ \citenamefont
  {Heller}(2003)}]{Vanicek_03}%
  \BibitemOpen
  \bibfield  {author} {\bibinfo {author} {\bibfnamefont {J.}~\bibnamefont
  {Van\'{\i}\v{c}ek}}\ and\ \bibinfo {author} {\bibfnamefont {E.~J.}\
  \bibnamefont {Heller}},\ }\href {\doibase 10.1103/PhysRevE.68.056208}
  {\bibfield  {journal} {\bibinfo  {journal} {Phys.\ Rev.~E}\ }\textbf
  {\bibinfo {volume} {68}},\ \bibinfo {pages} {056208} (\bibinfo {year}
  {2003})}\BibitemShut {NoStop}%
\bibitem [{\citenamefont {Miller}\ and\ \citenamefont
  {Smith}(1978)}]{MillerSmith_PRA_78}%
  \BibitemOpen
  \bibfield  {author} {\bibinfo {author} {\bibfnamefont {W.~H.}\ \bibnamefont
  {Miller}}\ and\ \bibinfo {author} {\bibfnamefont {F.~T.}\ \bibnamefont
  {Smith}},\ }\href {\doibase 10.1103/PhysRevA.17.939} {\bibfield  {journal}
  {\bibinfo  {journal} {Phys.\ Rev.~A}\ }\textbf {\bibinfo {volume} {17}},\
  \bibinfo {pages} {939} (\bibinfo {year} {1978})}\BibitemShut {NoStop}%
\bibitem [{\citenamefont {Hubbard}\ and\ \citenamefont
  {Miller}(1983)}]{Hubbard_JCP_83}%
  \BibitemOpen
  \bibfield  {author} {\bibinfo {author} {\bibfnamefont {L.~M.}\ \bibnamefont
  {Hubbard}}\ and\ \bibinfo {author} {\bibfnamefont {W.~H.}\ \bibnamefont
  {Miller}},\ }\href {\doibase 10.1063/1.444976} {\bibfield  {journal}
  {\bibinfo  {journal} {J.~Chem.\ Phys.}\ }\textbf {\bibinfo {volume} {78}},\
  \bibinfo {pages} {1801} (\bibinfo {year} {1983})}\BibitemShut {NoStop}%
\bibitem [{\citenamefont {Poulsen}, \citenamefont {Nyman},\ and\ \citenamefont
  {Rossky}(2003)}]{Poulsen_03}%
  \BibitemOpen
  \bibfield  {author} {\bibinfo {author} {\bibfnamefont {J.~A.}\ \bibnamefont
  {Poulsen}}, \bibinfo {author} {\bibfnamefont {G.}~\bibnamefont {Nyman}}, \
  and\ \bibinfo {author} {\bibfnamefont {P.~J.}\ \bibnamefont {Rossky}},\
  }\href {\doibase 10.1063/1.1626631} {\bibfield  {journal} {\bibinfo
  {journal} {J.~Chem.\ Phys.}\ }\textbf {\bibinfo {volume} {119}},\ \bibinfo
  {pages} {12179} (\bibinfo {year} {2003})}\BibitemShut {NoStop}%
\bibitem [{\citenamefont {Bonella}\ and\ \citenamefont
  {Coker}(2005)}]{Bonella_05}%
  \BibitemOpen
  \bibfield  {author} {\bibinfo {author} {\bibfnamefont {S.}~\bibnamefont
  {Bonella}}\ and\ \bibinfo {author} {\bibfnamefont {D.~F.}\ \bibnamefont
  {Coker}},\ }\href {\doibase 10.1063/1.1896948} {\bibfield  {journal}
  {\bibinfo  {journal} {J.~Chem.\ Phys.}\ }\textbf {\bibinfo {volume} {122}},\
  \bibinfo {eid} {194102} (\bibinfo {year} {2005})}\BibitemShut {NoStop}%
\bibitem [{\citenamefont {Huo}\ and\ \citenamefont {Coker}(2011)}]{Coker_11}%
  \BibitemOpen
  \bibfield  {author} {\bibinfo {author} {\bibfnamefont {P.}~\bibnamefont
  {Huo}}\ and\ \bibinfo {author} {\bibfnamefont {D.~F.}\ \bibnamefont
  {Coker}},\ }\href {\doibase 10.1063/1.3664763} {\bibfield  {journal}
  {\bibinfo  {journal} {J.~Chem.\ Phys.}\ }\textbf {\bibinfo {volume} {135}},\
  \bibinfo {eid} {201101} (\bibinfo {year} {2011})}\BibitemShut {NoStop}%
\bibitem [{\citenamefont {Heller}(1975)}]{Heller_75}%
  \BibitemOpen
  \bibfield  {author} {\bibinfo {author} {\bibfnamefont {E.~J.}\ \bibnamefont
  {Heller}},\ }\href {\doibase 10.1063/1.430620} {\bibfield  {journal}
  {\bibinfo  {journal} {J.~Chem.\ Phys.}\ }\textbf {\bibinfo {volume} {62}},\
  \bibinfo {pages} {1544} (\bibinfo {year} {1975})}\BibitemShut {NoStop}%
\bibitem [{\citenamefont {Davis}\ and\ \citenamefont
  {Heller}(1979)}]{Davis_79}%
  \BibitemOpen
  \bibfield  {author} {\bibinfo {author} {\bibfnamefont {M.~J.}\ \bibnamefont
  {Davis}}\ and\ \bibinfo {author} {\bibfnamefont {E.~J.}\ \bibnamefont
  {Heller}},\ }\href {\doibase 10.1063/1.438727} {\bibfield  {journal}
  {\bibinfo  {journal} {J.~Chem.\ Phys.}\ }\textbf {\bibinfo {volume} {71}},\
  \bibinfo {pages} {3383} (\bibinfo {year} {1979})}\BibitemShut {NoStop}%
\bibitem [{\citenamefont {Heller}(1981)}]{Heller_81}%
  \BibitemOpen
  \bibfield  {author} {\bibinfo {author} {\bibfnamefont {E.~J.}\ \bibnamefont
  {Heller}},\ }\href {\doibase 10.1063/1.442382} {\bibfield  {journal}
  {\bibinfo  {journal} {J.~Chem.\ Phys.}\ }\textbf {\bibinfo {volume} {75}},\
  \bibinfo {pages} {2923} (\bibinfo {year} {1981})}\BibitemShut {NoStop}%
\bibitem [{\citenamefont {Beck}\ \emph {et~al.}(2000)\citenamefont {Beck},
  \citenamefont {J\"{a}ckle}, \citenamefont {Worth},\ and\ \citenamefont
  {Meyer}}]{Beck_00}%
  \BibitemOpen
  \bibfield  {author} {\bibinfo {author} {\bibfnamefont {M.}~\bibnamefont
  {Beck}}, \bibinfo {author} {\bibfnamefont {A.}~\bibnamefont {J\"{a}ckle}},
  \bibinfo {author} {\bibfnamefont {G.}~\bibnamefont {Worth}}, \ and\ \bibinfo
  {author} {\bibfnamefont {H.-D.}\ \bibnamefont {Meyer}},\ }\href {\doibase
  10.1016/S0370-1573(99)00047-2} {\bibfield  {journal} {\bibinfo  {journal}
  {Phys.~Rep.}\ }\textbf {\bibinfo {volume} {324}},\ \bibinfo {pages} {1}
  (\bibinfo {year} {2000})}\BibitemShut {NoStop}%
\bibitem [{\citenamefont {Heller}(2006)}]{Heller_06}%
  \BibitemOpen
  \bibfield  {author} {\bibinfo {author} {\bibfnamefont {E.~J.}\ \bibnamefont
  {Heller}},\ }\href {\doibase 10.1021/ar040196y} {\bibfield  {journal}
  {\bibinfo  {journal} {Acc.~Chem.~Res.}\ }\textbf {\bibinfo {volume} {39}},\
  \bibinfo {pages} {127} (\bibinfo {year} {2006})}\BibitemShut {NoStop}%
\bibitem [{\citenamefont {Heller}(1976)}]{Heller_JCP_76}%
  \BibitemOpen
  \bibfield  {author} {\bibinfo {author} {\bibfnamefont {E.~J.}\ \bibnamefont
  {Heller}},\ }\href {\doibase 10.1063/1.431911} {\bibfield  {journal}
  {\bibinfo  {journal} {J.~Chem.\ Phys.}\ }\textbf {\bibinfo {volume} {64}},\
  \bibinfo {pages} {63} (\bibinfo {year} {1976})}\BibitemShut {NoStop}%
\bibitem [{\citenamefont {Skodje}\ and\ \citenamefont
  {Truhlar}(1984)}]{Truhlar_JCP_84}%
  \BibitemOpen
  \bibfield  {author} {\bibinfo {author} {\bibfnamefont {R.~T.}\ \bibnamefont
  {Skodje}}\ and\ \bibinfo {author} {\bibfnamefont {D.~G.}\ \bibnamefont
  {Truhlar}},\ }\href {\doibase 10.1063/1.447127} {\bibfield  {journal}
  {\bibinfo  {journal} {J.~Chem.\ Phys.}\ }\textbf {\bibinfo {volume} {80}},\
  \bibinfo {pages} {3123} (\bibinfo {year} {1984})}\BibitemShut {NoStop}%
\bibitem [{note1()}]{note1}%
  \BibitemOpen
  \bibinfo {note} {Gaussian state $|\phi_{j,\alpha}(t;\,\tau)\rangle$ is
  obtained in two consecutive steps: the initial frozen Gaussian
  $|\phi_{\alpha}\rangle$ is first propagated on the excited surface for time
  $\tau$ and then on the $j$th surface for time $t$. States
  $|\phi_{0,\alpha}(0;\,\tau)\rangle$ and $|\phi^1_{\alpha}(0;\,\tau)\rangle$
  are therefore identical.}\BibitemShut {Stop}%
\bibitem [{\citenamefont {Shalashilin}(2010)}]{Shalashilin:2010}%
  \BibitemOpen
  \bibfield  {author} {\bibinfo {author} {\bibfnamefont {D.~V.}\ \bibnamefont
  {Shalashilin}},\ }\href {\doibase 10.1063/1.3442747} {\bibfield  {journal}
  {\bibinfo  {journal} {J.~Chem.\ Phys.}\ }\textbf {\bibinfo {volume} {132}},\
  \bibinfo {eid} {244111} (\bibinfo {year} {2010})}\BibitemShut {NoStop}%
\bibitem [{\citenamefont {Tatchen}\ \emph {et~al.}(2011)\citenamefont
  {Tatchen}, \citenamefont {Pollak}, \citenamefont {Tao},\ and\ \citenamefont
  {Miller}}]{Tatchen_11}%
  \BibitemOpen
  \bibfield  {author} {\bibinfo {author} {\bibfnamefont {J.}~\bibnamefont
  {Tatchen}}, \bibinfo {author} {\bibfnamefont {E.}~\bibnamefont {Pollak}},
  \bibinfo {author} {\bibfnamefont {G.}~\bibnamefont {Tao}}, \ and\ \bibinfo
  {author} {\bibfnamefont {W.~H.}\ \bibnamefont {Miller}},\ }\href {\doibase
  10.1063/1.3573566} {\bibfield  {journal} {\bibinfo  {journal} {J.~Chem.\
  Phys.}\ }\textbf {\bibinfo {volume} {134}},\ \bibinfo {eid} {134104}
  (\bibinfo {year} {2011})}\BibitemShut {NoStop}%
\bibitem [{\citenamefont {Lee}\ and\ \citenamefont {Heller}(1982)}]{Lee_82}%
  \BibitemOpen
  \bibfield  {author} {\bibinfo {author} {\bibfnamefont {S.-Y.}\ \bibnamefont
  {Lee}}\ and\ \bibinfo {author} {\bibfnamefont {E.~J.}\ \bibnamefont
  {Heller}},\ }\href {\doibase 10.1063/1.443342} {\bibfield  {journal}
  {\bibinfo  {journal} {J.~Chem.\ Phys.}\ }\textbf {\bibinfo {volume} {76}},\
  \bibinfo {pages} {3035} (\bibinfo {year} {1982})}\BibitemShut {NoStop}%
\bibitem [{\citenamefont {Stock}\ \emph {et~al.}(1995)\citenamefont {Stock},
  \citenamefont {Woywod}, \citenamefont {Domcke}, \citenamefont {Swinney},\
  and\ \citenamefont {Hudson}}]{Stock_95}%
  \BibitemOpen
  \bibfield  {author} {\bibinfo {author} {\bibfnamefont {G.}~\bibnamefont
  {Stock}}, \bibinfo {author} {\bibfnamefont {C.}~\bibnamefont {Woywod}},
  \bibinfo {author} {\bibfnamefont {W.}~\bibnamefont {Domcke}}, \bibinfo
  {author} {\bibfnamefont {T.}~\bibnamefont {Swinney}}, \ and\ \bibinfo
  {author} {\bibfnamefont {B.~S.}\ \bibnamefont {Hudson}},\ }\href {\doibase
  10.1063/1.470689} {\bibfield  {journal} {\bibinfo  {journal} {J.~Chem.\
  Phys.}\ }\textbf {\bibinfo {volume} {103}},\ \bibinfo {pages} {6851}
  (\bibinfo {year} {1995})}\BibitemShut {NoStop}%
\bibitem [{\citenamefont {Duschinsky}(1937)}]{Duschinsky_1937}%
  \BibitemOpen
  \bibfield  {author} {\bibinfo {author} {\bibfnamefont {F.}~\bibnamefont
  {Duschinsky}},\ }\href@noop {} {\bibfield  {journal} {\bibinfo  {journal}
  {Acta Physicochim. URSS}\ }\textbf {\bibinfo {volume} {7}},\ \bibinfo {pages}
  {551} (\bibinfo {year} {1937})}\BibitemShut {NoStop}%
\bibitem [{\citenamefont {Meyer}(1986)}]{Meyer_86}%
  \BibitemOpen
  \bibfield  {author} {\bibinfo {author} {\bibfnamefont {H.-D.}\ \bibnamefont
  {Meyer}},\ }\href {\doibase 10.1063/1.450296} {\bibfield  {journal} {\bibinfo
   {journal} {J.~Chem.\ Phys.}\ }\textbf {\bibinfo {volume} {84}},\ \bibinfo
  {pages} {3147} (\bibinfo {year} {1986})}\BibitemShut {NoStop}%
\bibitem [{\citenamefont {Waterland}\ \emph {et~al.}(1988)\citenamefont
  {Waterland}, \citenamefont {Yuan}, \citenamefont {Martens}, \citenamefont
  {Gillilan},\ and\ \citenamefont {Reinhardt}}]{Waterland_88}%
  \BibitemOpen
  \bibfield  {author} {\bibinfo {author} {\bibfnamefont {R.~L.}\ \bibnamefont
  {Waterland}}, \bibinfo {author} {\bibfnamefont {J.-M.}\ \bibnamefont {Yuan}},
  \bibinfo {author} {\bibfnamefont {C.~C.}\ \bibnamefont {Martens}}, \bibinfo
  {author} {\bibfnamefont {R.~E.}\ \bibnamefont {Gillilan}}, \ and\ \bibinfo
  {author} {\bibfnamefont {W.~P.}\ \bibnamefont {Reinhardt}},\ }\href {\doibase
  10.1103/PhysRevLett.61.2733} {\bibfield  {journal} {\bibinfo  {journal}
  {Phys.\ Rev.\ Lett.}\ }\textbf {\bibinfo {volume} {61}},\ \bibinfo {pages}
  {2733} (\bibinfo {year} {1988})}\BibitemShut {NoStop}%
\bibitem [{\citenamefont {Eckhardt}, \citenamefont {Hose},\ and\ \citenamefont
  {Pollak}(1989)}]{Pollak_89}%
  \BibitemOpen
  \bibfield  {author} {\bibinfo {author} {\bibfnamefont {B.}~\bibnamefont
  {Eckhardt}}, \bibinfo {author} {\bibfnamefont {G.}~\bibnamefont {Hose}}, \
  and\ \bibinfo {author} {\bibfnamefont {E.}~\bibnamefont {Pollak}},\ }\href
  {\doibase 10.1103/PhysRevA.39.3776} {\bibfield  {journal} {\bibinfo
  {journal} {Phys.\ Rev.~A}\ }\textbf {\bibinfo {volume} {39}},\ \bibinfo
  {pages} {3776} (\bibinfo {year} {1989})}\BibitemShut {NoStop}%
\bibitem [{\citenamefont {Bohigas}, \citenamefont {Tomsovic},\ and\
  \citenamefont {Ullmo}(1993)}]{Bohigas_93}%
  \BibitemOpen
  \bibfield  {author} {\bibinfo {author} {\bibfnamefont {O.}~\bibnamefont
  {Bohigas}}, \bibinfo {author} {\bibfnamefont {S.}~\bibnamefont {Tomsovic}}, \
  and\ \bibinfo {author} {\bibfnamefont {D.}~\bibnamefont {Ullmo}},\ }\href
  {\doibase 10.1016/0370-1573(93)90109-Q} {\bibfield  {journal} {\bibinfo
  {journal} {Phys.~Rep.}\ }\textbf {\bibinfo {volume} {223}},\ \bibinfo {pages}
  {43} (\bibinfo {year} {1993})}\BibitemShut {NoStop}%
\bibitem [{\citenamefont {Revuelta}\ \emph {et~al.}(2012)\citenamefont
  {Revuelta}, \citenamefont {Vergini}, \citenamefont {Benito},\ and\
  \citenamefont {Borondo}}]{Revuelta_12}%
  \BibitemOpen
  \bibfield  {author} {\bibinfo {author} {\bibfnamefont {F.}~\bibnamefont
  {Revuelta}}, \bibinfo {author} {\bibfnamefont {E.~G.}\ \bibnamefont
  {Vergini}}, \bibinfo {author} {\bibfnamefont {R.~M.}\ \bibnamefont {Benito}},
  \ and\ \bibinfo {author} {\bibfnamefont {F.}~\bibnamefont {Borondo}},\ }\href
  {\doibase 10.1103/PhysRevE.85.026214} {\bibfield  {journal} {\bibinfo
  {journal} {Phys.\ Rev.~E}\ }\textbf {\bibinfo {volume} {85}},\ \bibinfo
  {pages} {026214} (\bibinfo {year} {2012})}\BibitemShut {NoStop}%
\bibitem [{\citenamefont {Zambrano}, \citenamefont {\v{S}ulc},\ and\
  \citenamefont {Van\'{i}\v{c}ek}()}]{Zambrano_tba}%
  \BibitemOpen
  \bibfield  {author} {\bibinfo {author} {\bibfnamefont {E.}~\bibnamefont
  {Zambrano}}, \bibinfo {author} {\bibfnamefont {M.}~\bibnamefont {\v{S}ulc}},
  \ and\ \bibinfo {author} {\bibfnamefont {J.}~\bibnamefont
  {Van\'{i}\v{c}ek}},\ }\href@noop {} {}\bibinfo {note} {In
  preparation}\BibitemShut {NoStop}%
\bibitem [{\citenamefont {Shalashilin}\ and\ \citenamefont
  {Child}(2008)}]{DVS_pancakes}%
  \BibitemOpen
  \bibfield  {author} {\bibinfo {author} {\bibfnamefont {D.~V.}\ \bibnamefont
  {Shalashilin}}\ and\ \bibinfo {author} {\bibfnamefont {M.~S.}\ \bibnamefont
  {Child}},\ }\href {\doibase 10.1063/1.2828509} {\bibfield  {journal}
  {\bibinfo  {journal} {J.~Chem.\ Phys.}\ }\textbf {\bibinfo {volume} {128}},\
  \bibinfo {eid} {054102} (\bibinfo {year} {2008})}\BibitemShut {NoStop}%
\bibitem [{\citenamefont {Gradshteyn}\ and\ \citenamefont
  {Ryzhik}(2007)}]{book_ryzhik}%
  \BibitemOpen
  \bibfield  {author} {\bibinfo {author} {\bibfnamefont {I.~S.}\ \bibnamefont
  {Gradshteyn}}\ and\ \bibinfo {author} {\bibfnamefont {I.~M.}\ \bibnamefont
  {Ryzhik}},\ }\href@noop {} {\emph {\bibinfo {title} {{Table of Integrals,
  Series, and Products}}}},\ \bibinfo {edition} {7th}\ ed.\ (\bibinfo
  {publisher} {Academic Press},\ \bibinfo {address} {San Diego},\ \bibinfo
  {year} {2007})\BibitemShut {NoStop}%
\bibitem [{\citenamefont {Brewer}, \citenamefont {Hulme},\ and\ \citenamefont
  {Manolopoulos}(1997)}]{Brewer_97}%
  \BibitemOpen
  \bibfield  {author} {\bibinfo {author} {\bibfnamefont {M.~L.}\ \bibnamefont
  {Brewer}}, \bibinfo {author} {\bibfnamefont {J.~S.}\ \bibnamefont {Hulme}}, \
  and\ \bibinfo {author} {\bibfnamefont {D.~E.}\ \bibnamefont {Manolopoulos}},\
  }\href {\doibase 10.1063/1.473532} {\bibfield  {journal} {\bibinfo  {journal}
  {J.~Chem.\ Phys.}\ }\textbf {\bibinfo {volume} {106}},\ \bibinfo {pages}
  {4832} (\bibinfo {year} {1997})}\BibitemShut {NoStop}%
\bibitem [{\citenamefont {Koonin}(1998)}]{book_Koonin}%
  \BibitemOpen
  \bibfield  {author} {\bibinfo {author} {\bibfnamefont {S.}~\bibnamefont
  {Koonin}},\ }\href@noop {} {\emph {\bibinfo {title} {Computational Physics:
  Fortran version}}}\ (\bibinfo  {publisher} {Westview Press},\ \bibinfo {year}
  {1998})\BibitemShut {NoStop}%
\bibitem [{\citenamefont {Press}\ \emph {et~al.}(2007)\citenamefont {Press},
  \citenamefont {Teukolsky}, \citenamefont {Vetterling},\ and\ \citenamefont
  {Flannery}}]{book_NR}%
  \BibitemOpen
  \bibfield  {author} {\bibinfo {author} {\bibfnamefont {W.~H.}\ \bibnamefont
  {Press}}, \bibinfo {author} {\bibfnamefont {S.~A.}\ \bibnamefont
  {Teukolsky}}, \bibinfo {author} {\bibfnamefont {W.~T.}\ \bibnamefont
  {Vetterling}}, \ and\ \bibinfo {author} {\bibfnamefont {B.~P.}\ \bibnamefont
  {Flannery}},\ }\href@noop {} {\emph {\bibinfo {title} {Numerical Recipes, The
  art of scientific computing}}},\ \bibinfo {edition} {3rd}\ ed.\ (\bibinfo
  {publisher} {Cambridge University Press},\ \bibinfo {year}
  {2007})\BibitemShut {NoStop}%
\bibitem [{\citenamefont {Levine}\ \emph {et~al.}(2008)\citenamefont {Levine},
  \citenamefont {Coe}, \citenamefont {Virshup},\ and\ \citenamefont
  {Mart\'{\i}nez}}]{Levine:2008}%
  \BibitemOpen
  \bibfield  {author} {\bibinfo {author} {\bibfnamefont {B.~G.}\ \bibnamefont
  {Levine}}, \bibinfo {author} {\bibfnamefont {J.~D.}\ \bibnamefont {Coe}},
  \bibinfo {author} {\bibfnamefont {A.~M.}\ \bibnamefont {Virshup}}, \ and\
  \bibinfo {author} {\bibfnamefont {T.~J.}\ \bibnamefont {Mart\'{\i}nez}},\
  }\href {\doibase 10.1016/j.chemphys.2008.01.014} {\bibfield  {journal}
  {\bibinfo  {journal} {Chem.\ Phys.}\ }\textbf {\bibinfo {volume} {347}},\
  \bibinfo {pages} {3 } (\bibinfo {year} {2008})}\BibitemShut {NoStop}%
\bibitem [{\citenamefont {Kay}(1989)}]{Kay_89}%
  \BibitemOpen
  \bibfield  {author} {\bibinfo {author} {\bibfnamefont {K.~G.}\ \bibnamefont
  {Kay}},\ }\href {\doibase 10.1016/0301-0104(89)87102-2} {\bibfield  {journal}
  {\bibinfo  {journal} {Chem.\ Phys.}\ }\textbf {\bibinfo {volume} {137}},\
  \bibinfo {pages} {165} (\bibinfo {year} {1989})}\BibitemShut {NoStop}%
\bibitem [{\citenamefont {Andersson}(2001)}]{Andersson_01}%
  \BibitemOpen
  \bibfield  {author} {\bibinfo {author} {\bibfnamefont {L.~M.}\ \bibnamefont
  {Andersson}},\ }\href {\doibase 10.1063/1.1380204} {\bibfield  {journal}
  {\bibinfo  {journal} {J.~Chem.\ Phys.}\ }\textbf {\bibinfo {volume} {115}},\
  \bibinfo {pages} {1158} (\bibinfo {year} {2001})}\BibitemShut {NoStop}%
\bibitem [{\citenamefont {Sidje}(1998)}]{EXPOKIT}%
  \BibitemOpen
  \bibfield  {author} {\bibinfo {author} {\bibfnamefont {R.}~\bibnamefont
  {Sidje}},\ }\href {http://www.expokit.org} {\bibfield  {journal} {\bibinfo
  {journal} {ACM Trans.\ Math.\ Softw.}\ }\textbf {\bibinfo {volume} {24}},\
  \bibinfo {pages} {130} (\bibinfo {year} {1998})}\BibitemShut {NoStop}%
\bibitem [{\citenamefont {Burant}\ and\ \citenamefont
  {Batista}(2002)}]{Burant_02}%
  \BibitemOpen
  \bibfield  {author} {\bibinfo {author} {\bibfnamefont {J.~C.}\ \bibnamefont
  {Burant}}\ and\ \bibinfo {author} {\bibfnamefont {V.~S.}\ \bibnamefont
  {Batista}},\ }\href {\doibase 10.1063/1.1436306} {\bibfield  {journal}
  {\bibinfo  {journal} {J.~Chem.\ Phys.}\ }\textbf {\bibinfo {volume} {116}},\
  \bibinfo {pages} {2748} (\bibinfo {year} {2002})}\BibitemShut {NoStop}%
\bibitem [{\citenamefont {Kay}(1994)}]{Kay_94c}%
  \BibitemOpen
  \bibfield  {author} {\bibinfo {author} {\bibfnamefont {K.~G.}\ \bibnamefont
  {Kay}},\ }\href {\doibase 10.1063/1.466273} {\bibfield  {journal} {\bibinfo
  {journal} {J.~Chem.\ Phys.}\ }\textbf {\bibinfo {volume} {100}},\ \bibinfo
  {pages} {4432} (\bibinfo {year} {1994})}\BibitemShut {NoStop}%
\bibitem [{\citenamefont {Ben-Nun}, \citenamefont {Quenneville},\ and\
  \citenamefont {Mart\'{\i}nez}(2000)}]{Martinez_JPCA_00}%
  \BibitemOpen
  \bibfield  {author} {\bibinfo {author} {\bibfnamefont {M.}~\bibnamefont
  {Ben-Nun}}, \bibinfo {author} {\bibfnamefont {J.}~\bibnamefont
  {Quenneville}}, \ and\ \bibinfo {author} {\bibfnamefont {T.~J.}\ \bibnamefont
  {Mart\'{\i}nez}},\ }\href {\doibase 10.1021/jp994174i} {\bibfield  {journal}
  {\bibinfo  {journal} {J.~Phys.\ Chem.~A}\ }\textbf {\bibinfo {volume}
  {104}},\ \bibinfo {pages} {5161} (\bibinfo {year} {2000})}\BibitemShut
  {NoStop}%
\bibitem [{\citenamefont {Wu}\ and\ \citenamefont {Batista}(2003)}]{Wu_03}%
  \BibitemOpen
  \bibfield  {author} {\bibinfo {author} {\bibfnamefont {Y.}~\bibnamefont
  {Wu}}\ and\ \bibinfo {author} {\bibfnamefont {V.~S.}\ \bibnamefont
  {Batista}},\ }\href {\doibase 10.1063/1.1560636} {\bibfield  {journal}
  {\bibinfo  {journal} {J.~Chem.\ Phys.}\ }\textbf {\bibinfo {volume} {118}},\
  \bibinfo {pages} {6720} (\bibinfo {year} {2003})}\BibitemShut {NoStop}%
\end{thebibliography}

\end{document}